\documentclass[preprint,preprintnumbers,nofootinbib,tightenlines,12pt]{revtex4}

\usepackage{graphicx}
\usepackage{amssymb}
\usepackage{amsmath,mathrsfs,verbatim}
\usepackage{times}
\usepackage{latexsym}
\def\bey{\begin{eqnarray}}
\def\eey{\end{eqnarray}}

\def\lsim{\mathrel{\raise.3ex\hbox{$<$\kern-.75em\lower1ex\hbox{$\sim$}}}}
\def\gsim{\mathrel{\raise.3ex\hbox{$>$\kern-.75em\lower1ex\hbox{$\sim$}}}}

\newcommand{\mx}{m_X}
\newcommand{\ax}{\alpha_X}
\newcommand{\mphi}{m_\phi}

\usepackage{color}
\definecolor{rosso}{cmyk}{0,1,1,0.4}
\definecolor{rossoCP3}{cmyk}{0,.88,.77,.40}
\definecolor{verdeCP3}{rgb}{0.09765625, 0.57421875, 0.1015625}
\definecolor{bluCP3}{rgb}{0, 0.23, 0.67}

\newcommand{\eg}{e.g.~}
\newcommand{\ie}{i.e.~}

\newcommand{\Eq}[1]{Eq.~\eqref{#1}}
\newcommand{\Fig}[1]{Fig.~\ref{#1}}

\newcommand{\beq}{\begin{equation}}
\newcommand{\eeq}{\end{equation}}
\newcommand{\ud}{\text{d}}

\newcommand{\bol}[1]{\boldsymbol{#1}}

\newcommand{\ER}{E_\text{R}}
\newcommand{\Ed}{E'}
\newcommand{\vesc}{v_\text{esc}}
\newcommand{\vmin}{v_\text{min}}

\ifx\pdfoutput\undefined
\usepackage[dvips,bookmarks]{hyperref}    
\else
\usepackage{hyperref}    
\fi
\hypersetup{colorlinks, bookmarksopen, bookmarksnumbered,
citecolor=verdeCP3, linkcolor=bluCP3, pdfstartview=FitH, urlcolor=rossoCP3}

\usepackage{floatrow}
\newfloatcommand{capbtabbox}{figure}[][\FBwidth]

\begin{document}

\title{Direct Detection Signatures of \\ Self-Interacting Dark Matter with a Light Mediator}

\author{Eugenio Del Nobile$^{a}$, Manoj Kaplinghat$^{b}$, and Hai-Bo Yu$^{c}$
\vspace{5mm}\\
$^{a}$\normalsize\emph{Department of Physics and Astronomy, University of California, Los Angeles, CA, 90095, USA}  \vspace{1mm} \\ 
$^{b}$\normalsize\emph{Department of Physics and Astronomy, University of California, Irvine, CA, 92697, USA}  \vspace{1mm} \\ 
$^{c}$\normalsize\emph{Department of Physics and Astronomy, University of California, Riverside, CA, 92507, USA}
}

\date{\today}

\begin{abstract}

\noindent
Self-interacting dark matter (SIDM) is a simple and well-motivated scenario that could explain long-standing puzzles in structure formation on small scales. If the required self-interaction arises through a light mediator (with mass $\sim 10$ MeV) in the dark sector, this new particle must be unstable to avoid overclosing the universe. The decay of the light mediator could happen due to a weak coupling of the hidden and visible sectors, providing new signatures for direct detection experiments. The SIDM nuclear recoil spectrum is more peaked towards low energies compared to the usual case of contact interactions, because the mediator mass is comparable to the momentum transfer of nuclear recoils. We show that the SIDM signal could be distinguished from that of DM particles with contact interactions by considering the time-average energy spectrum in experiments employing different target materials, or the average and modulated spectra in a single experiment. Using current limits from LUX and SuperCDMS, we also derive strong bounds on the mixing parameter between hidden and visible sector.

\end{abstract}

\maketitle

\section{Introduction}

Dark matter (DM) accounts for $85\%$ of the matter in the universe~\cite{Ade:2013zuv}, but its particle physics nature remains elusive. In one of the well-studied models, DM candidates have a weak-scale mass and carry weak-scale interactions. In this weakly interacting massive particle (WIMP) paradigm, the interaction strength between DM and the standard model (SM) sector is also weak-scale and the resulting relic density for WIMPs could be consistent with, or smaller than, the observed cosmological DM density. This coincidence of weak-scale and relic density has motivated many direct, indirect, and collider searches for WIMPs leading to significant regions of the model parameter space being ruled out. Alternatively, axions motivated by the strong CP problem could have the right relic density to be all of the DM and present laboratory searches and limits from stellar evolution have ruled out large regions of parameter space. 

Both WIMPs and axions are consistent with the prevailing view in astrophysics that DM is a cold non-interacting particle. The cold DM paradigm has been thoroughly tested on large scales. On galactic and sub-galactic scales, however, the data strongly suggest that our ideas of structure formation are incomplete. In particular, the DM densities inferred in the central regions of DM-dominated galaxies are, on average, lower than expected from DM-only simulations~\cite{Rocha:2012jg,Weinberg:2013aya}. Given this situation and the lack of a consistent signal in WIMP searches, it is important to thoroughly study DM candidates other than WIMPs. A compelling possibility is that DM may strongly interact with itself~\cite{Spergel:1999mh} (self-interacting dark matter or SIDM). DM self-interactions can transfer energy from the hotter, outer region to the inner region of DM halos, thereby reducing the central density and providing a possible solution to the small-scale puzzles. Recent numerical simulations have shown that SIDM halos are consistent with observations if DM particles scatter with each other with a nuclear-scale cross section within halos~\cite{Vogelsberger:2012ku, Rocha:2012jg,Peter:2012jh,Zavala:2012us,Vogelsberger:2014pda,Buckley:2014hja,Elbert:2014bma}.

The key ingredient of the majority of SIDM models is the existence of a light mediator with mass $\lesssim 100$ MeV~\cite{Feng:2009hw,Tulin:2012wi,Tulin:2013teo,Cline:2013pca,Hochberg:2014kqa,Mambrini:2015nza,Boddy:2014yra}, which must decay in the early universe to avoid overclosure~\cite{Kaplinghat:2013yxa,Boddy:2014yra,Zhang:2015era}. Unless additional massless degrees of freedom are introduced, the mediator naturally decays to SM particles through a mixing portal between the two sectors, which may lead to SIDM signals in indirect and direct detection experiments~\cite{Kaplinghat:2013yxa}. For example, SIDM particles may annihilate to mediators, which subsequently decay to electrons. These energetic electrons could generate gamma-ray signals in environments such as the galactic center through inverse Compton scattering on starlight~\cite{Kaplinghat:2015gha}. It has also been shown that DM direct detection experiments are sensitive to the SIDM parameter space even if the mixing parameter between the two sectors is extremely feeble~\cite{Kaplinghat:2013yxa,Ko:2014nha,Kouvaris:2014uoa,Li:2014vza}.

An interesting feature of SIDM in direct detection is that DM interacts with nuclei with a long-range force, in contrast to a contact interaction in WIMP models. The usual tree-level $t$-channel DM-nucleus differential scattering cross section scales as $(q^2+\mphi^2)^{-2}$, where $q$ is the momentum transfer in nuclear recoils and $\mphi$ is the mediator mass. The typical mediator mass in SIDM models is comparable to the typical momentum transfer in direct detection experiments and hence the SIDM event spectrum is more peaked toward low recoil energies, compared to WIMP models where $\mphi \gg q$. This novel feature of SIDM signals in direct detection was noted in Ref.~\cite{Kaplinghat:2013yxa}, where an approximation method was used to reinterpret direct detection bounds for WIMPs to constrain SIDM models. The effect of light mediators on direct detection bounds was also discussed recently in Ref.~\cite{Li:2014vza} (see also~\cite{Fornengo:2011sz, Cherry:2014wia}), with a focus on the bound dependence on the mediator mass and the determination of the DM mass from experimental data. In this paper we investigate direct detection of DM with light mediators in detail, paying special attention to their scattering spectra. We first derive a strong upper bound on the mixing parameter between the SIDM and SM sectors. Then we study the SIDM event spectrum in the light of SuperCDMS, LUX, and DAMA, taking into account realistic efficiency and energy resolution of the detectors. Considering both the time-average and modulated event rates, we show that direct detection experiments can potentially distinguish SIDM from WIMPs.

In the remainder of this work, we first present a simple particle physics model for SIDM, and discuss basics of DM direct detection in Sec.~\ref{sec:direct detection}. Our results are presented in Sec.~\ref{sec:contraints}. Lastly, we conclude in Sec.~\ref{sec:conclusions}.

\section{Particle physics model and direct detection rate}
\label{sec:direct detection}

\subsection{Particle physics model}
We assume that the DM particle $X$, either a Dirac fermion or a complex scalar, interacts with the vector mediator $\phi$ of a dark $U(1)_X$ gauge interaction. In the non-relativistic limit, self-interactions between DM particles can be described by a Yukawa potential~\cite{Feng:2009hw,Buckley:2009in,Loeb:2010gj,Aarssen:2012fx,Bellazzini:2013foa,Petraki:2014uza,Wise:2014jva,Schutz:2014nka}
\beq\label{eq:yukawa}
V(r)=\pm\frac{\ax}{r}e^{-\mphi r} \ ,
\eeq
where $\ax \equiv g_{X}/(4\pi) $ is the fine structure constant in the dark sector and $\mphi$ is the mediator mass. We fix $\ax=0.01$, motivated by the value of the electromagnetic fine structure constant in the SM. We also focus on the case of asymmetric SIDM in which only DM $X$, and not its anti-particle, is present in DM halos. Hence, DM self-scattering is purely repulsive and the ``$+$" sign of the Yukawa interaction in Eq.~(\ref{eq:yukawa}) must be considered.

In general, the dark sector can couple to the SM through the kinetic mixing $\epsilon_\gamma \, \phi_{\mu\nu} F^{\mu\nu}$~\cite{Holdom:1985ag}, where $\epsilon_\gamma$ is the mixing parameter, and $\phi_{\mu\nu}$ and $F^{\mu\nu}$ are the field strength of the mediator $\phi$ and of the photon, respectively. The mixing induces a coupling of $\phi$ to SM fermions $f$ at ${\cal O}(\epsilon_\gamma)$ upon diagonalization: $\epsilon_\gamma e\sum_f Q_f \bar{f}\gamma^\mu f \phi_\mu$, where $Q_f$ denotes the electric charge (in units of $e$) of the SM fermions. In this case, direct detection signals of SIDM arise from DM-proton scattering via $\phi$ exchange. Our analysis can be easily generalized to other cases such as $\phi$-$Z$ mixing portal, or Higgs portal for a scalar mediator~\cite{Kaplinghat:2013yxa}. Notice however that all these models have similar phenomenology at direct detection experiments; the main difference being that for the kinetic mixing case the DM interacts dominantly with protons, for the $Z$-mixing case mostly with neutrons, and for the Higgs portal case equally with protons and neutrons.

The differential cross section for DM-nucleus scattering is~\cite{Fornengo:2011sz,Kaplinghat:2013yxa}
\beq\label{sigma_XT}
\frac{\ud\sigma_{XT}}{\ud q^2} = \frac{4\pi\alpha_\text{em}\ax\epsilon_\gamma^2 Z^2}{(q^2+m^2_\phi)^2}\frac{1}{v^2}F_T^2(q^2) \ ,
\eeq
where $\alpha_\text{em}=1/137$ is the SM fine structure constant, $Z$ is the number of protons in the nucleus, $q$ is the momentum transfer, $v$ is the speed of the DM particle in the nucleus rest frame and $F_T(q^2)$ is the nuclear form factor related to the charge density in the nucleus~\cite{Helm:1956zz, Lewin:1995rx}. The nuclear recoil energy $\ER$ is related to the momentum transfer and the nuclear mass $m_T$ by $q = \sqrt{2 m_T \ER}$.

\begin{table}
  \begin{tabular}{| l | c | r || r | r |}
    \hline
    Model & $\mx$ (GeV) & $\mphi$ (MeV) & $q_\text{Xe}$ (MeV) & $q_\text{Ge}$ (MeV) \\ \hline
     A & $1000$ & $3$ & $127$ & $74$ \\ \hline
     B &  $100$ & $15$ & $62$ & $46$ \\ \hline
     C &  $10$ & $20$ & $10$ & $10$ \\ \hline
     D &  $5$ & $20$ & $5$ & $5$ \\
 \hline
   \end{tabular}
     \caption{SIDM benchmark models considered in this paper. In the two rightmost columns we indicate the typical values of the momentum transfer for recoils off xenon (relevant for LUX) and germanium (relevant for SuperCDMS) of a DM particle with typical speed in Earth's frame $v_\odot = 232$ km/s. The maximum attainable momentum transfer, assuming a maximum DM speed $v_\text{max} = \vesc + v_\odot = 776$ km/s, is $4.7$ times higher than this typical value, while values that are lower than those shown here are always possible. The average sensitivity windows of the two experiments are $[28 \text{ MeV}, 81 \text{ MeV}]$ for LUX and $[15 \text{ MeV}, 38 \text{ MeV}]$ for SuperCDMS (see footnote \ref{nu^-1}).}
  \label{tb:models}
  \end{table}

To investigate the signal spectrum of SIDM in SIDM-nucleus scattering, we choose four benchmark models as shown in Table~\ref{tb:models}. Also shown in Table~\ref{tb:models} are the typical values of the momentum transfer for recoils off xenon (relevant for LUX) and germanium (relevant for SuperCDMS) of a DM particle with typical speed in Earth's frame $v_\odot = 232$ km/s, $q \approx \sqrt{2} \mu_T v_\odot$ with $\mu_T$ the DM-nucleus reduced mass. While smaller values of the momentum transfer are always possible depending on the scattering angle, the maximum attainable momentum transfer is $2 \mu_T v_\text{max}$, with $v_\text{max}$ the maximum possible DM speed in Earth's frame (discussed later). We can anticipate for which models the long-range nature of the interaction will be important by comparing $m_\phi$ with the typical $q$.

\begin{figure}
\includegraphics[width=0.49\textwidth]{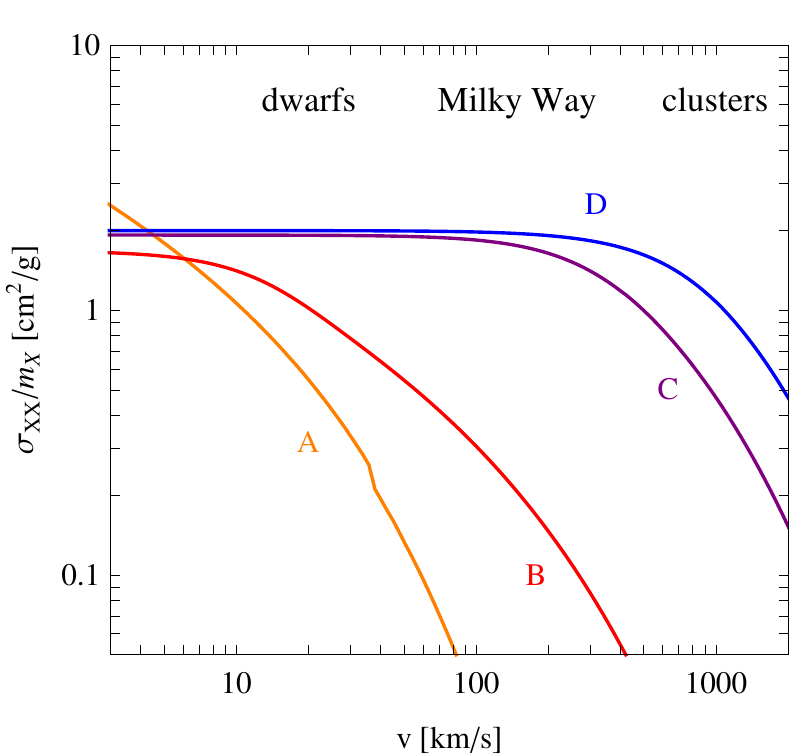}
\caption{DM self-scattering cross section per unit mass, as a function of the DM relative speed. Repulsive self-interaction is assumed. Curves correspond to the SIDM models A, B, C, D summarized in Table \ref{tb:models}.}
\label{fig:sigmaxx}
\end{figure}

Fig.~\ref{fig:sigmaxx} shows the DM self-scattering cross section of our four benchmark models as a function of the DM relative speed. In model A, the self-scattering cross section is suppressed significantly at large velocities, because DM self-scattering occurs in the Rutherford limit with $\sigma_{XX}\propto 1 / v^4$ on large scales. For model B, DM self-interactions are important in dwarf galaxies, and mildly in Milky Way-sized galaxies, but are suppressed on cluster scales. On the other hand, $\sigma_{XX} / \mx$ is relevant from dwarf to cluster scales for both model C and D.

\subsection{Scattering rate in direct detection experiments}
The differential recoil rate for a DM particle scattering off a target nucleus $T$, expressed in counts per day per kilogram per keV, is
\beq\label{RT}
\frac{\ud R_T}{\ud \ER}(t) = \frac{\xi_T}{m_T} \frac{\rho}{m_X} \int_{v \geqslant \vmin(\ER)} \hspace{-.60cm} \ud^3 v \, v \, f(\bol{v}, t) \frac{\ud \sigma_{XT}}{\ud \ER} (v, \ER) \ ,
\eeq
where $\xi_T / m_T$ is the number of targets per detector mass in units of kg$^{-1}$, $f(\bol{v}, t)$ is the DM velocity distribution in Earth's frame and $\rho$ is the local DM density, which we set to $0.3$ GeV/cm$^3$. $\vmin(\ER)$ is the minimum speed a DM particle must have to impart a recoil energy $\ER$ to the target nucleus; for elastic scattering,
\beq\label{eq:vmin}
\vmin(\ER) = \sqrt{m_T \ER / 2 \mu_T^2} \ .
\eeq

While no information is available at present on the DM velocity distribution in our galaxy, we notice that DM self-interactions would thermalize the halo, and therefore it is justified to consider a thermal velocity distribution~\cite{Vogelsberger:2012sa}. We assume the Standard Halo Model (SHM) for the DM halo, \ie an isothermal sphere with an isotropic Maxwellian velocity distribution in the galactic frame:
\beq
f_\text{G}({\bol{v}}) = \frac{e^{- v^2 / v_0^2}}{(v_0 \sqrt{\pi})^3 N_\text{esc}} \, \theta(\vesc - v) \ ,
\rule[-12pt]{0pt}{20pt}
\eeq
with $N_\text{esc} \equiv \text{erf}(v_\text{esc} / v_0) - 2 (v_\text{esc} / v_0) \exp(- v_\text{esc}^2 / v_0^2) / \sqrt{\pi}$ a normalization factor that ensures that $\int \ud^3 v \, f_\text{G}({\bol{v}}) = 1$. The distribution is truncated at the galactic escape speed $\vesc = 544$ km/s~\cite{Smith:2006ym}, and we set $v_0$ equal to the speed of the Local Standard of Rest, $v_0 = 220$ km/s. The DM velocity distribution $f(\bol{v}, t)$ in Earth's frame is related to $f_\text{G}(\bol{v})$ by the Galilean transformation $f(\bol{v}, t) = f_\text{G}(\bol{v} + \bol{v}_\text{obs}(t))$, with $\bol{v}_\text{obs}(t)$ the velocity of Earth with respect to the galactic frame. This is $\bol{v}_\text{obs}(t) = \bol{v}_\odot + \bol{v}_\oplus(t)$, where $\bol{v}_\oplus(t)$ is Earth's rotational velocity around the Sun and $\bol{v}_\odot$ is the Sun's velocity with respect to the galactic frame. The average Earth's speed over a revolution period of one year equals $| \langle \bol{v}_\text{obs} \rangle | = v_\odot = 232$ km/s. Notice that, because of the Galilean boost, the maximum speed attainable in Earth's frame is (in average) $v_\text{max} = \vesc + v_\odot$.

Given the dependency of the differential cross section in \Eq{sigma_XT} on the DM speed, $\ud \sigma_{XT} / \ud \ER \propto 1 / v^2$, which is common to many non-relativistic DM-nucleus scattering processes, the relevant velocity integral entering the differential rate in \Eq{RT} is
\beq\label{eq:eta}
\eta(\vmin, t) \equiv \int_{v \geqslant \vmin}
\hspace{-.20cm} 
\ud^3 v \, \frac{f(\bol{v}, t)}{v} \ .
\eeq
Notice that the DM velocity distribution being normalized implies that also $\eta(\vmin, t)$ is normalized in $\vmin$-space,
\beq\label{etanormalization}
\int_0^\infty \ud\vmin \, \eta(\vmin, t) = 1 \ ,
\eeq
at any time $t$. This can be seen by denoting $F(v, t) \equiv \int \ud\Omega_{\bol{v}} \, v^2 f(\bol{v}, t)$ with $\ud^3 v = v^2 \ud v \, \ud\Omega_{\bol{v}}$, so that
$
\int_0^\infty \ud\vmin \, \eta(\vmin, t) =
\int_0^\infty \ud\vmin \, \int_{\vmin}^\infty \ud v \, F(v, t) / v
$,
and then inverting the integration order.

If the DM halo is smooth and isotropic,\footnote{Anisotropies and non-smooth components in the DM halo can be due to DM substructure, see \eg\cite{Lee:2013xxa, DelNobile:2015nua}, and to the Sun's gravitational focusing of DM particles~\cite{Lee:2013wza, Bozorgnia:2014dqa, DelNobile:2015nua}. These can noticeably modify the average rate and the modulation of the signal with respect to those of an isotropic halo.} since $v_\odot, v_0 \gg v_\oplus = 30$ km/s we can Taylor-expand the rate in \Eq{RT} in the dimensionless parameter
\beq
z(t) \equiv \frac{| \bol{v}_\text{obs}(t) |}{v_0} \simeq \frac{v_\odot}{v_0} + b \cos[\omega (t - t_0)] \frac{v_\oplus}{v_0}
\eeq
about its average value $v_\odot / v_0$, where $t_0$ is the time when $\bol{v}_\oplus(t)$ and $\bol{v}_\odot$ are aligned, $\omega = 2 \pi /$yr and $b \approx 0.5$ (see \eg\cite{Freese:2012xd}). Therefore we have for the velocity integral in \Eq{eq:eta}
\beq\label{etaexpansion}
\eta(\vmin, t) \simeq \underbrace{\eta(\vmin, t) \bigg|_{z(t) = \frac{v_\odot}{v_0}}}_{\displaystyle \equiv \eta_0(\vmin)} + \underbrace{b \frac{v_\oplus}{v_0} \, \frac{\ud}{\ud z} \eta(\vmin, t) \bigg|_{z(t) = \frac{v_\odot}{v_0}}}_{\displaystyle \equiv \eta_1(\vmin)} \cos[\omega (t - t_0)] \ ,
\eeq
where we identified the unmodulated (time-average) and modulated components of the velocity integral, $\eta_0$ and $\eta_1$ respectively (notice that these are time-independent quantities). While $\eta_0$ is positive by definition, $\eta_1$ can also assume negative values. \Eq{etanormalization} implies
\begin{align}\label{eta01normalization}
\int_0^\infty \ud\vmin \, \eta_0(\vmin) = 1 \ ,
&&&
\int_0^\infty \ud\vmin \, \eta_1(\vmin) = 0 \ ,
\end{align}
as can be inferred by taking $t = t_0 + \frac{\pi}{2}$. These equalities hold as long as the approximation in \Eq{etaexpansion} is valid. \Eq{etaexpansion} implies that the differential rate depends on time approximately as
\beq\label{rateexpansion}
\frac{\ud R_T}{\ud \ER}(t) \simeq \frac{\ud R_T}{\ud \ER} + \frac{\ud R_T^\text{mod}}{\ud \ER} \cos[\omega (t - t_0)] \ .
\eeq
The factor multiplying $\cos[\omega (t - t_0)]$ is the modulated (component of the) rate, while the first term is the average or unmodulated rate, for which we keep the same symbol as the total rate, with a slight abuse of notation, since $\ud R_T^\text{mod} / \ud \ER$ is negligible in comparison. As with $\eta_0$ and $\eta_1$, the modulated and the unmodulated rates are time-independent quantities, and $\ud R_T^\text{mod} / \ud \ER$ can assume negative values.

To actually compare the theoretical rate in \Eq{RT} with the event rate measured by the experiments, we now have to take into account detection efficiency, cuts acceptance and energy resolution of the detectors. We consider LUX and SuperCDMS for the measure of the unmodulated rate, and DAMA for the measure of the modulated rate.
 
For LUX, we determine the differential rate in $S1$ (primary scintillation light, measured in photoelectrons phe) following the procedure used by the similar experiment XENON100~\cite{Aprile:2011hx}: the number $n$ of photons created by a recoiling Xe nucleus in the liquid phase, which is Poisson distributed, is smeared at detection due to the finite resolution of the photomultiplier tubes (PMT) whose effect is parameterized by a Gaussian distribution. The measured differential rate is then
\beq\label{diffrateLUX}
\frac{\ud R}{\ud S1} = \epsilon(S1) \sum_{n = 1}^\infty {\rm Gauss}(S1 | n, \sqrt n \, \sigma_{\rm PMT}) \int_{3 \, {\rm keV}}^\infty \ud \ER \, {\rm Poiss}(n | \nu(\ER)) \sum_T \frac{\ud R_T}{\ud \ER} \ ,
\eeq
where $\sigma_{\rm PMT} = 0.37$ is the average single photoelectron resolution of the PMT's~\cite{Akerib:2012ys}, and $T$ denotes the different Xe isotopes. ${\rm Gauss}(x | \mu, \sigma)$ and ${\rm Poiss}(x | \mu)$ denote respectively Gaussian and Poisson probability distributions for $x$ with mean $\mu$ and, for the Gaussian, standard deviation $\sigma$. For the efficiency $\epsilon(S1)$ we take the dashed red line in the top panel of Fig.~1 of Ref.~\cite{Akerib:2013tjd}. We set a conservative lower threshold of $3$ keV on the modeled signal~\cite{Akerib:2013tjd}. $\nu(\ER)$, the average number of photons generated by a Xe nucleus recoiling with energy $\ER$, is obtained from the spreadsheet available on the NEST website~\cite{NEST} using as input $2.888$ g/cm$^3$ for the density of liquid xenon and $181$ V/cm for the drift field~\cite{SzydagisPC, Szydagis:2014xog}. The energy range where the signal $S1$ is measured is $2$--$30$ phe.

For SuperCDMS, the measured differential recoil rate in detected energy $\Ed$ is
\beq\label{diffrateSuperCDMS}
\frac{\ud R}{\ud \Ed} = \int_0^\infty \ud\ER \, \epsilon(\ER) {\rm Gauss}(\Ed | \ER, \sigma(\Ed)) \sum_T \frac{\ud R_T}{\ud \ER} \ ,
\eeq
with the detector resolution $\sigma(\Ed) = \sqrt{0.293^2 + 0.056^2 \Ed / \text{keV}}$ adopted for the CDMS-II germanium detectors~\cite{Ahmed:2009rh}, and the efficiency $\epsilon(\ER)$ taken to be the red line in Fig.~1 of Ref.~\cite{Agnese:2014aze}. $T$ here denotes the different Ge isotopes. The energy range where the signal $\Ed$ is measured is $1.6$--$10$ keV.

We also study the annual modulation signal of SIDM, taking the DAMA experiment as an example~\cite{Bernabei:2013xsa}. Our study can also be applied straightforwardly to future DAMA-like experiments as KIMS-NaI, ANAIS, DM-Ice17, and SABRE (see \eg\cite{Cooley:2014aya} and references therein). DAMA collected events in the energy range $2$--$20$ keV for an exposure of $1.33$ ton$\, \cdot \,$yr. The Collaboration measures the modulated rate without attempting any background subtraction. The measured modulated rate is
\beq
\frac{\ud R_\text{mod}}{\ud \Ed} = \int_0^\infty \ud\ER \sum_T {\rm Gauss}(\Ed | Q_T \ER, \sigma(Q_T \ER)) \frac{\ud R_T^\text{mod}}{\ud \ER} \ ,
\eeq
with the detector resolution $\sigma(E) = 0.448 \sqrt{E} + 0.0091 E$~\cite{Bernabei:2008yh}. The sum over target nuclei counts sodium and iodine, with quenching factors $Q_{\rm Na} = 0.3$ and $Q_{\rm I} = 0.09$ respectively.

\section{Direct detection of self-interacting dark matter}
\label{sec:contraints}

\subsection{Current constraints from LUX and SuperCDMS}
In the first LUX results~\cite{Akerib:2013tjd}, the collected data were found consistent with the background-only hypothesis. With an exposure of $\omega = 85.3$ day$\, \cdot \, 118.3$ kg, one single candidate event was found below the mean of the nuclear recoil distribution in the $S1$--$\log_{10}(S2 / S1)$ plane (solid red line in Figs.~3 and 4 of Ref.~\cite{Akerib:2013tjd}), where 0.64 background events were estimated. We consider the DM signal in the same region by assuming a $50\%$ efficiency for DM events below the nuclear recoil mean for heavy enough DM particles. We therefore compute the number of signal events as $0.50 \times w \int_2^{30} \ud S1 \, \frac{\ud R}{\ud S1}$.
To set a constraint on the kinetic mixing parameter $\epsilon_\gamma$ for SIDM models, we require the p-value of the data to be smaller than the significance level $\alpha = 10\%$, assuming a Poisson distribution for the total number of events. This leads to an upper limit of $3.25$ events on the total number of signal events in LUX.

In SuperCDMS, a blind analysis with an exposure $577$ kg$\,\cdot\,$day revealed 11 candidate events against an estimated background of $6.1$ events~\cite{Agnese:2014aze}. We derive an upper bound on $\epsilon_\gamma$ as above, yielding an upper bound of $10.5$ signal events in SuperCDMS.

In deriving our bounds, we only use the total number of events in the signal range instead of using the full spectral information of the rate. While a spectral analysis is ultimately needed in order to distinguish SIDM from WIMPs (see next section), extracting information on the particle physics model of DM interactions relies on assuming a model for the DM distribution. The SHM, used in this work, although motivated in the framework of SIDM, is only a first guess. The effect of baryons, anisotropies in the DM distribution and DM substructure will affect the velocity distribution. Our bounds are therefore somewhat solid in this respect, in the sense that they allow for small variations of the DM distribution, that can modify the detected event spectrum, but such that the total number of events remains fixed.

\begin{figure}
\centering
\includegraphics[width=0.49\textwidth]{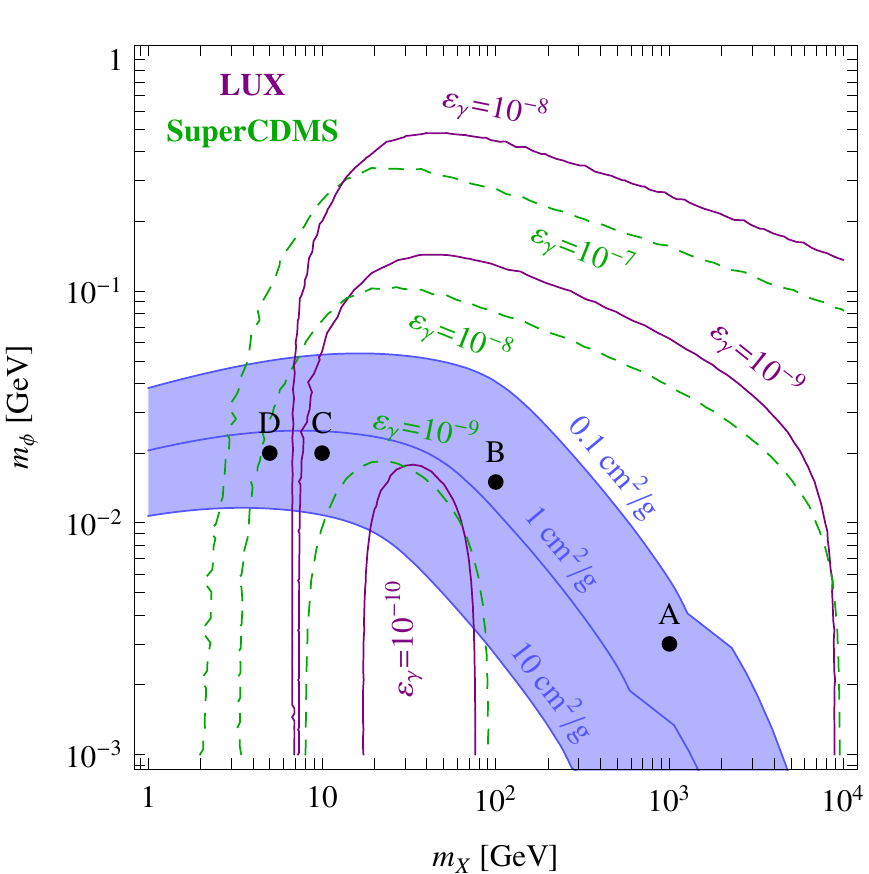}
\includegraphics[width=0.49\textwidth]{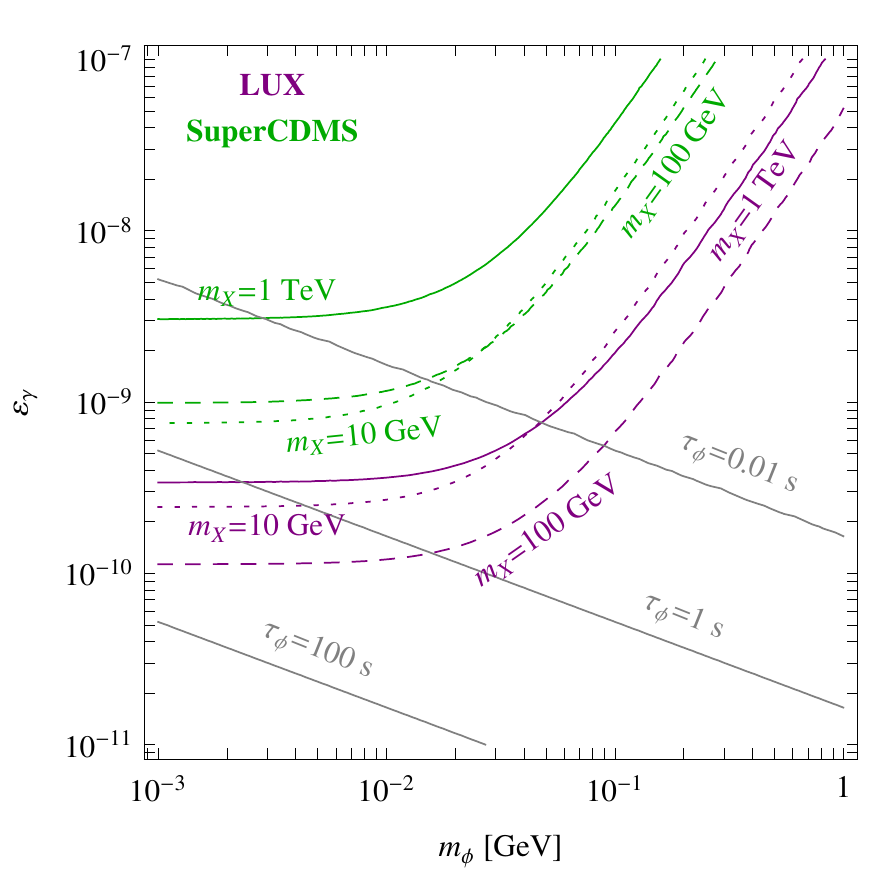}
\caption{Direct detection constraints on SIDM parameter space, assuming $\ax=0.01$. {\it Left:} Lower limits on the $(m_X, m_\phi)$ plane, for different values of $\epsilon_\gamma$, from LUX (purple lines) and SuperCDMS (dashed green lines). The region below each curve is excluded at a significance level $1 - \alpha = 90\%$. The shaded band is where SIDM solves structure anomalies on dwarf scales. {\it Right:} Upper limits on the $(m_\phi, \epsilon_\gamma)$ plane, for different values of $m_X$. The region above each curve is excluded at a significance level $1 - \alpha = 90\%$. The gray lines are curves of constant decay time $\tau_\phi = 0.01$, $1$, and $100$ s for the mediator $\phi$.}
\label{fig:constraints}
\end{figure}

Fig.~\ref{fig:constraints} shows direct detection constraints on the SIDM parameter space. The blue lines in the left panel denote the portion of parameter space where SIDM could explain the small scale anomalies, for three different values of the self-interaction cross section per unit mass $\sigma_{XX} / \mx = 0.1, 1, 10 \text{ cm}^2 / \text{g}$. It is clear that both LUX and SuperCDMS put a strong constraint on the mixing parameter $\epsilon_\gamma$. As can be seen in the left panel of Fig.~\ref{fig:constraints}, LUX excludes all favored $(\mx,\mphi)$ regions with $\epsilon_\gamma \gtrsim 10^{-9}$ except for $\mx \lesssim 7$ GeV. The SuperCDMS limit is weaker, but it can exclude SIDM models with $\mx > 3$ GeV. Remarkably, benchmark points A, B and C are ruled out by LUX for $\epsilon_\gamma = 10^{-9}$, while benchmark point D can not be excluded by LUX because of the small DM mass (see below). It can however be excluded by SuperCDMS for $\epsilon_\gamma \sim 10^{-8}$ due to its lower energy threshold and its lighter target compared to xenon. It is remarkable that the LUX and SuperCDMS constraints on $\epsilon_\gamma$ are much stronger than those from beam dump experiments~\cite{Essig:2013lka}, which exclude for instance $\epsilon_\gamma \gtrsim 3 \times 10^{-8}$ for $m_\phi \lesssim 100$ MeV. Thus, direct detection experiments provide a unique window for exploring the dark sector. 

Fig.~\ref{fig:constraints} ({\it right}) shows the exclusion region in the $(\mphi, \epsilon_\gamma)$ plane for given $\mx$. For $\mphi \lesssim 10 \text{ MeV}$, the upper bound on $\epsilon_\gamma$ becomes nearly independent of $\mphi$ because the typical momentum transfer is much larger than the mediator mass, $q \gg \mphi$. On the other hand, when $\mphi \gg 10$ MeV, the bound follows the scaling relation $\epsilon_\gamma / \mphi^2 =$ constant for fixed DM mass, since in the limit of contact interaction $\mphi \gg q$ the direct detection cross section in \Eq{sigma_XT} scales as $\epsilon_\gamma^2 / \mphi^4$. The gray lines in the plot are curves of constant mediator lifetime, for the three values $\tau_\phi = 0.01, 1, 100$ s. Here $\tau_\phi = 1 / \Gamma_\phi$, with the decay rate $\Gamma_\phi = \alpha_\text{em} m_\phi \epsilon_\gamma^2 / 3$ dominated by decays into $e^+ e^-$~\cite{Kaplinghat:2013yxa}. If the mediator decays before weak freeze-out, $\sim 1$ s, then we can be assured that the predicted light element abundances will not be modified from the standard scenario. From Fig.~\ref{fig:constraints} ({\it right}) we see that direct detection experiments are probing the SIDM parameter space which is relevant to the thermal history of the early universe.

To further narrow down the allowed parameter range for $\epsilon_\gamma \lesssim 10^{-10}$, a detailed study of the impact of decays on Big Bang Nucleosynthesis (BBN) is required. It has been shown that BBN provides a model-independent constraint on $\epsilon_\gamma$ between $10^{-14}$ and $10^{-12}$, due to the destruction of $^3$He and $^2$H by the electromagnetic showers resulting from $\phi$ decays, with the conservative assumption that mediator particles are produced in the early universe by inverse decays only~\cite{Fradette:2014sza}. This limit needs to be reevaluated for the case where the dark photons were thermally populated in the thermal bath of the hidden sector~\cite{Feng:2008mu,Feng:2009mn}.

Since the range $10^{-12} \lesssim \epsilon_\gamma \lesssim 10^{-10}$  will be tested by direct detection experiments soon, let us consider BBN constraints on $\epsilon_\gamma$ in this range in some more detail. If the photons in the shower can scatter off background photons and produce electron-position pairs, then the distribution of photons in the shower is rapidly pushed to energies below $\sim m_e^2 / (22 T_\gamma) = 0.012 (\text{MeV} / T_\gamma)$ MeV~\cite{Kawasaki:1994sc,Holtmann:1998gd}, as long as the mediator mass is sufficiently larger than this threshold. As an example, this implies that the bulk of energy in the shower created by decays at $T_\gamma=100$ keV ($\tau_\phi \sim 100$ s) will be shifted rapidly to energies below the $^2$H binding energy, $2.2$ MeV. There could, however, be an effect on the light element abundances because the presence of the thermalized mediator (or its decay products) modifies the expansion rate during BBN. The magnitude of this effect depends on the temperature of the hidden sector, which could be very different from that of the visible sector since kinetic mixing with the low $\epsilon_\gamma$ values under consideration could not have kept the two sectors in thermal equilibrium~\cite{Feng:2010zp}. Based on these arguments, there should be regions of SIDM parameter space with $10^{-12} \lesssim \epsilon_\gamma \lesssim 10^{-10}$ which are consistent with all the BBN constraints and accessible to direct detection experiments.

It is also suggestive to compare direct detection bounds with other astrophysical constraints. For mediator masses in the range of interest here, $m_\phi \lesssim 100$ MeV, the parameter space with $\epsilon_\gamma \gtrsim 10^{-10}$ is strongly disfavored by supernova cooling arguments (similar to those for axions)~\cite{Dreiner:2013mua,Dent:2012mx,Bjorken:2009mm}. This constraint becomes ineffective when the coupling is large enough to trap the mediator particles within the supernova core, which happens above $\epsilon_\gamma \simeq 10^{-7}$. Interestingly, the lower limit from the supernova cooling constraint will be superseded by direct detection experiments soon. In fact, for $\mx\sim100$ GeV, the current LUX bound, $\epsilon_\gamma\lesssim10^{-10}$, is already comparable to the supernova cooling limit.

\medskip

\begin{figure}
\centering
\includegraphics[width=0.49\textwidth]{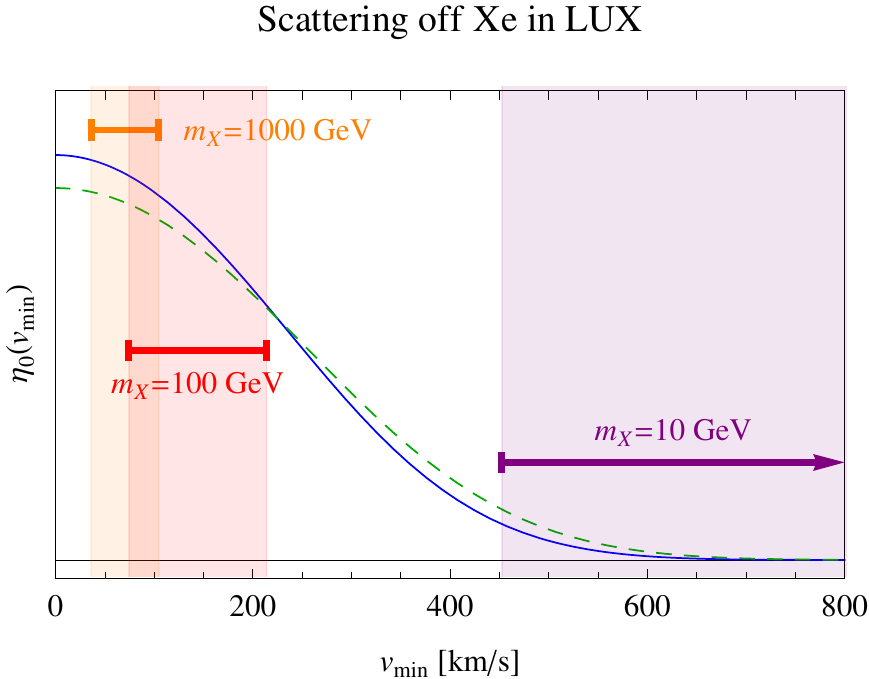}
\includegraphics[width=0.49\textwidth]{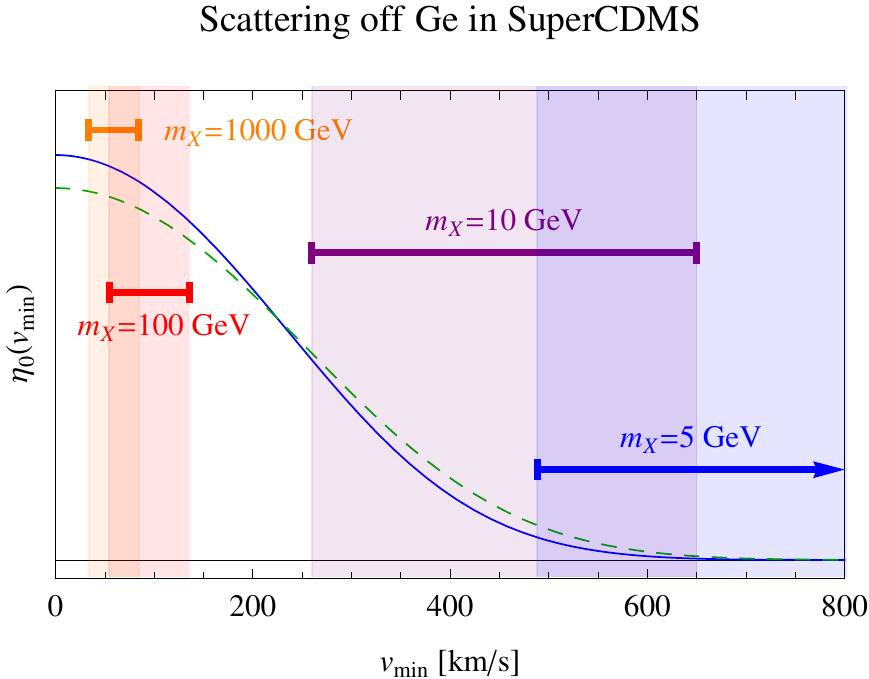}

\vspace{3mm}

\includegraphics[width=0.49\textwidth]{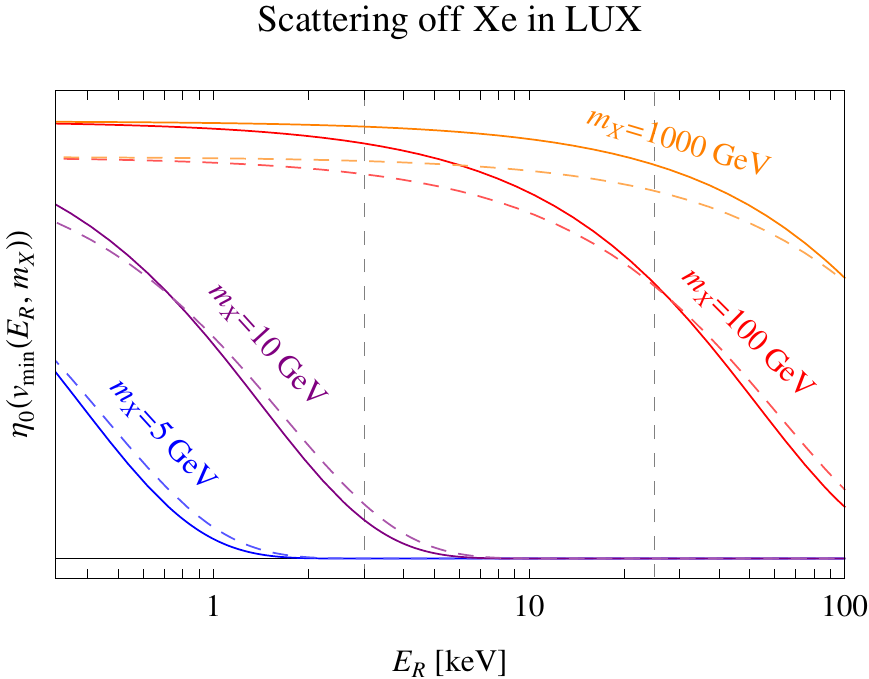}
\includegraphics[width=0.49\textwidth]{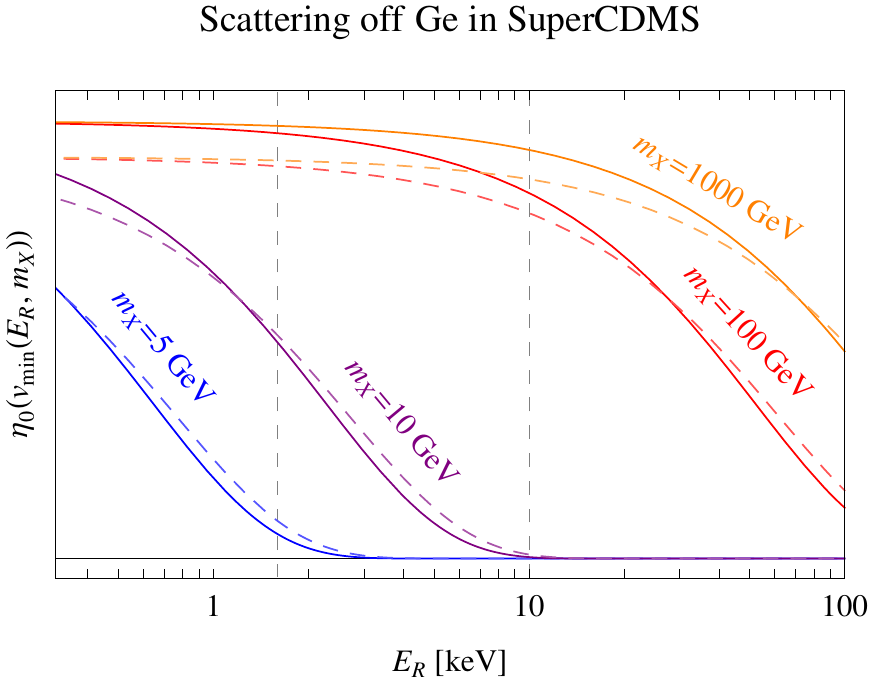}
\caption{Velocity integral in the SHM, with arbitrary units: $\eta_0(\vmin)$ in $\vmin$-space ({\it top}) and $\eta_0(\vmin(\ER, m_X))$ in $\ER$-space ({\it bottom}), for both LUX ({\it left}) and SuperCDMS ({\it right}). In all the plots, the solid lines are the velocity integral for our standard parameter values $v_0 = 220$ km/s, $v_\odot = 232$ km/s and $\vesc = 544$ km/s, while the dashed lines are for the larger values $v_0 = 240$ km/s, $v_\odot = 252$ km/s and $\vesc = 600$ km/s. The velocity integral is unique in $\vmin$-space, while it depends on the DM and target mass in $\ER$-space. On the other hand, recoil energies depend on $m_X$ and $m_T$ when shown in $\vmin$-space. The dashed vertical lines in the bottom panels show the indicative recoil energy range where the experiments are sensitive: $[3 \text{ keV}, 25 \text{ keV}]$ for DM scattering off xenon in LUX and $[1.6 \text{ keV}, 10 \text{ keV}]$ for scattering off germanium in SuperCDMS. The same ranges are shown in $\vmin$-space in the upper panels as bands of different colors for different values of $m_X$; notice that the band for $m_X = 10$ GeV in LUX and that for $m_X = 5$ GeV in SuperCDMS continue to higher values of $\vmin$ than those plotted, while the band for $m_X = 5$ GeV in LUX lies entirely outside of the plotted range.}
\label{fig:eta}
\end{figure}

We now focus our attention back on the bounds in the $(m_X, m_\phi)$ plane. To get a better understanding of the bounds, it is useful to indicatively write the total average rate in a certain recoil energy interval $[{\ER}_1, {\ER}_2]$ as
\beq\label{simple rate}
R(m_X, m_\phi, \epsilon_\gamma) \sim \int_{{\ER}_1}^{{\ER}_2} \ud \ER \, \frac{\epsilon_\gamma^2 \, \eta_0(\vmin(\ER, m_X))}{m_X (2 m_T \ER + m_\phi^2)^2} \ .
\eeq
Here we neglected all numerical constants and normalization factors, apart from $\epsilon_\gamma$, since we are only interested in the functional dependence of the rate on the model parameters, and we made explicit the dependence of $\vmin$ on $m_X$. We also neglected the nuclear form factor as it has no bearing on the following discussion. The presence of multiple isotopes in the detector and the signal smearing due to the finite detector resolution should not affect our argument, either. $\eta_0(\vmin)$ is plotted, for the SHM, in the top panels of \Fig{fig:eta} for two different sets of parameter values: our standard choice $v_0 = 220$ km/s, $v_\odot = 232$ km/s and $\vesc = 544$ km/s (solid blue line), and the slightly larger values $v_0 = 240$ km/s, $v_\odot = 252$ km/s and $\vesc = 600$ km/s (dashed green line). As expected, larger values of $v_0$ cause the velocity integral $\eta_0(\vmin)$ to extend to larger values of the minimum speed, thus $\eta_0$ is enhanced at large $\vmin$. Since $\int_0^\infty \ud \vmin \, \eta_0(\vmin) = 1$ (see \Eq{eta01normalization}), this also implies a reduction of the velocity integral for small $\vmin$. In the bottom panels of \Fig{fig:eta} we show again the velocity integral but now in $\ER$-space, $\eta_0(\vmin(\ER, m_X))$, for our four benchmark values of the DM mass, $m_X = 5$, $10$, $100$, and $1000$ GeV. For each value of the DM mass, the solid line corresponds to our standard set of SHM parameters while the dashed line is for the larger values reported above. The variables nuclear recoil energy $\ER$ and minimum speed $\vmin$ can be used interchangeably, and one can switch from one to the other via the $m_X$-dependent change of variables \Eq{eq:vmin}. The plots in $\vmin$-space and in $\ER$-space in \Fig{fig:eta} show the same thing in different manners, and we think it is useful to present them together.

Due to the correspondence between $\ER$ and $\vmin$, one can think of the rate in \Eq{simple rate} as the integral in $\ud \vmin$ of $\eta_0(\vmin)$ times some function of $\vmin$. The extrema of this integral depend on ${\ER}_1$ and ${\ER}_2$ and, crucially, on the DM mass. To illustrate this point, we show in \Fig{fig:eta} the approximate recoil energy ranges where LUX ({\it left}) and SuperCDMS ({\it right}) are sensitive,\footnote{\label{nu^-1} For LUX we adopt $[{\ER}_1, {\ER}_2] = [\nu^{-1}(2 \text{ phe}), \nu^{-1}(30 \text{ phe})] \simeq [3 \text{ keV}, 25 \text{ keV}]$, where $\nu^{-1}(S1)$ is the inverse function of the average photon number function $\nu(\ER)$ introduced earlier. The recoil energy range shown for SuperCDMS is $[1.6 \text{ keV}, 10 \text{ keV}]$. Note that these ranges should only be used for qualitative analysis, as a signal outside of these intervals may still be detected due to statistical fluctuations.} both in $\vmin$-space (colored shaded bands, {\it top}) for our four $m_X$ benchmark values, and in $\ER$-space (vertical dashed gray lines, {\it bottom}). It is clear that, while heavy DM samples small $\vmin$ values, light DM samples the high-velocity tail of the distribution: this is expected, since light DM particles need higher speeds to impart a certain recoil energy to the nucleus, compared to heavier DM particles. For small enough $m_X$, the experiment stops being sensitive to DM scattering events because there are no DM particles in the halo that are energetic enough to scatter a target nucleus above the experimental threshold. In other words, the $\vmin$ range probed by the experiment lies completely above the maximum possible DM speed in Earth's reference frame, $v_\text{max}$, and the only way of gaining sensitivity to such light DM particles is lowering the experimental energy threshold. This is the reason why a $5$ GeV DM particle as in our benchmark model D lies outside of LUX's reach (with the $3$ keV threshold), while being within the reach of SuperCDMS which employs a lighter target and has a lower energy threshold (see the left panel of Fig.~\ref{fig:constraints}).

In the opposite regime of heavy DM, or more quantitatively $m_X \gg m_T$, the minimum speed in \Eq{eq:vmin} becomes independent of $m_X$ and therefore the $\vmin$ range probed by the experiment does not change when increasing the DM mass. Thus, an experiment like LUX probes the same range in $\vmin$-space for a $1000$ GeV DM particle as for heavier DM particles, and the same happens with SuperCDMS. For this reason, the rate \Eq{simple rate} at these two experiments simply scales as $R \sim 1 / m_X$ for $m_X > 1000$ GeV.

We are now ready to understand the dependence of the limits shown in \Fig{fig:constraints} on the model parameters. In the simplest case where an experimental result is employed to only set an upper limit $R_\text{limit}$ on the predicted rate, a bound on the model parameters is set by requiring $R_\text{limit} > R(m_X, m_\phi, \epsilon_\gamma)$. Recalling what we just said about the velocity integral $\eta_0(\vmin)$, let us first notice that, for $\epsilon_\gamma$ and $m_\phi$ fixed, the rate in \Eq{simple rate} approaches zero for $m_X \to \infty$, and vanishes for all DM masses small enough that $\vmin(\ER, m_X) > v_\text{max}$ for all values of $\ER$ probed by the experiment. For intermediate values of the DM mass, the rate is non-zero and it has therefore a global maximum, which is reached in the massless mediator limit (effectively attained in the long-range regime $m_\phi^2 \ll 2 m_T \ER$). Therefore, there exist a value of the mixing parameter $\epsilon_\gamma$ below which no bounds can be set, because for this value the rate is smaller than $R_\text{limit}$ in all of the parameter space.

Above this critical $\epsilon_\gamma$, bounds can be determined in the following way. Starting from a point in parameter space where $m_X$ is very large and $m_\phi = 0$ (bottom-right portion of the left panel of \Fig{fig:constraints}), we can decrease $m_X$ keeping the mixing parameter fixed until we obtain $R = R_\text{limit}$. Models with DM mass larger than that so determined are viable, while those with similar but smaller $m_X$ predict a larger rate and are therefore ruled out by the experiment. For fixed $\epsilon_\gamma$, the limit line in parameter space can be determined by progressively `turning on' $m_\phi$; the limit line raises vertical in the $(m_X, m_\phi)$ plane until $m_\phi$ grows of the same order of magnitude as the momentum transfer $2 m_T \ER$, when the rate becomes sensitive to the mediator mass and a raise in $m_\phi$ can be compensated by decreasing $m_X$, so that the predicted rate \Eq{simple rate} remains equal to the limit rate. We then enter the contact-interaction regime, where $m_\phi \gg 2 m_T \ER$ and, if the DM mass is still large enough that $\eta_0(\vmin(\ER, m_X))$ changes slowly with decreasing $m_X$ in the $\ER$ range probed by the experiment, the limit line scales in the parameter space as $\epsilon_\gamma^2 / (m_X m_\phi^4) =$ constant, as can be seen from \Eq{simple rate}. If we keep following the limit line for smaller and smaller $m_X$ values, eventually the DM mass becomes so small that the velocity integral vanishes quickly because all scattering events occur below the experimental threshold. Such a reduction in the rate can be compensated by a decrease in $m_\phi$, if we keep $\epsilon_\gamma$ fixed, until we enter again the long-range regime where the rate is no longer sensitive to $m_\phi$ and the limit line drops vertically in the $(m_X, m_\phi)$ plane. For smaller values of the mixing parameter, however, the condition $R = R_\text{limit}$ can be obtained even before reaching the kinematical limit of the experiment.

We can now also easily understand the effect of changing the halo parameters or the experimental threshold on the parameter space limits. \Fig{fig:eta} shows that increasing the typical DM speed makes $\eta_0(\vmin)$ larger at large $\vmin$ and smaller at low $\vmin$ values, \ie the differential rate increases at high energies and decreases at low energies. Consequently, the total rate in \Eq{simple rate} will be smaller for heavy DM, which probes small values of $\vmin$, while it will be larger for light DM. For fixed $\epsilon_\gamma$ and $m_\phi \ll 2 m_T \ER$, this change can be contrasted by a decrease in $m_X$, since $\eta_0(\vmin(\ER, m_X)) / m_X$ increases (decreases) with the DM mass for light (heavy) enough DM at fixed $\ER$, so to maintain the limit condition $R = R_\text{limit}$. Therefore, a larger DM average speed will cause the limits in \Fig{fig:constraints} to move slightly to smaller values of $m_X$. Decreasing the experimental energy threshold has the general effect of increasing the rate, thus the limits on the parameter space will become more stringent. The bounds on the $(m_X, m_\phi)$ plane will therefore move to larger $m_X$ for heavy DM and to smaller $m_X$ for light DM.

\subsection{Identifying SIDM with LUX and SuperCDMS}
To see whether DM direct detection experiments are able to distinguish SIDM candidates from usual WIMPs if a positive signal is detected, we perform a careful study of the SIDM signal spectrum in SuperCDMS and LUX. Given the current constraints on the mixing parameter as discussed above, we take $\epsilon_\gamma = 10^{-10}$ for the SIDM models as a baseline. We note that the BBN constraints discussed earlier may require $\epsilon_\gamma \gtrsim 10^{-12}$ and therefore current and next generation experiments should be able to test much of the viable parameter space of this simple SIDM setup. For each of our four SIDM benchmark models, we also consider two additional cases, for comparison: one is with the mediator mass to be three times the value of $\mphi$ in the benchmark model; the other is a WIMP model, where a spin-independent (SI) contact interaction between DM and nuclei is assumed. The SI cross section can be taken to be that in \Eq{sigma_XT} in the limit $m_\phi \gg q$, and with arbitrary values of the coupling constants. We normalize the coupling strength of the two additional models such that they have the same total number of events (proportional to the integral of the measured rate) as the benchmark model. The model with three times larger mediator mass is meant to provide a comparison point between the benchmark SIDM model (exhibiting long-range dynamics due to the light mediator) and the contact interaction of the SI model. 

\begin{figure}[t!]
\centering
\includegraphics[width=0.49\textwidth]{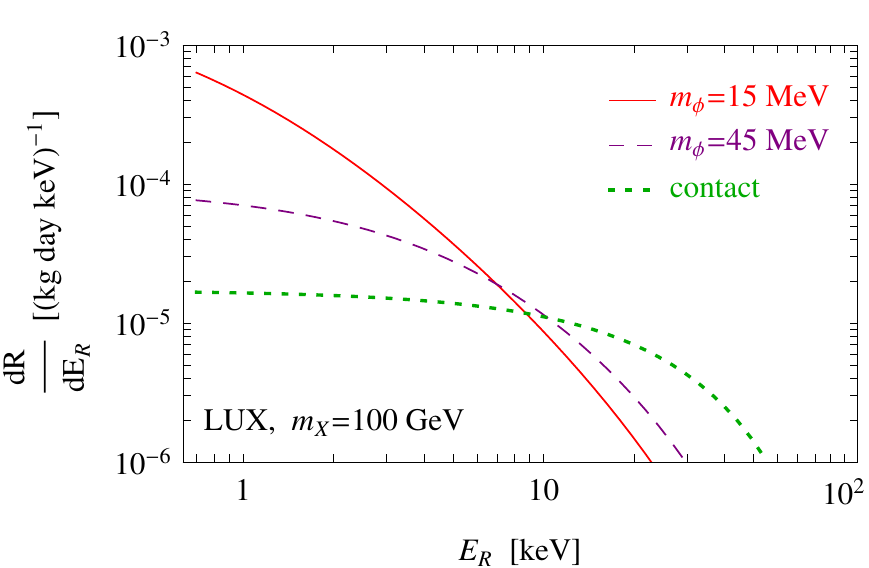}
\includegraphics[width=0.49\textwidth]{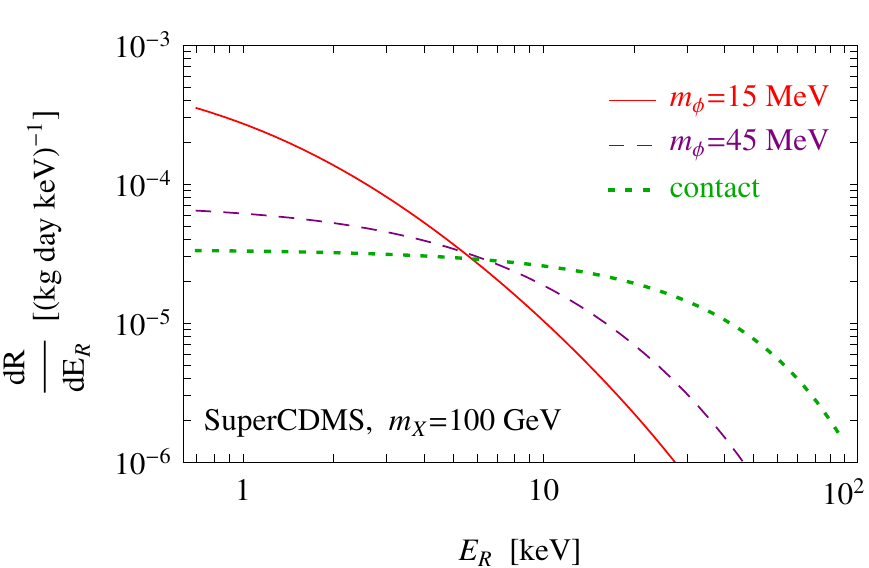}
\includegraphics[width=0.49\textwidth]{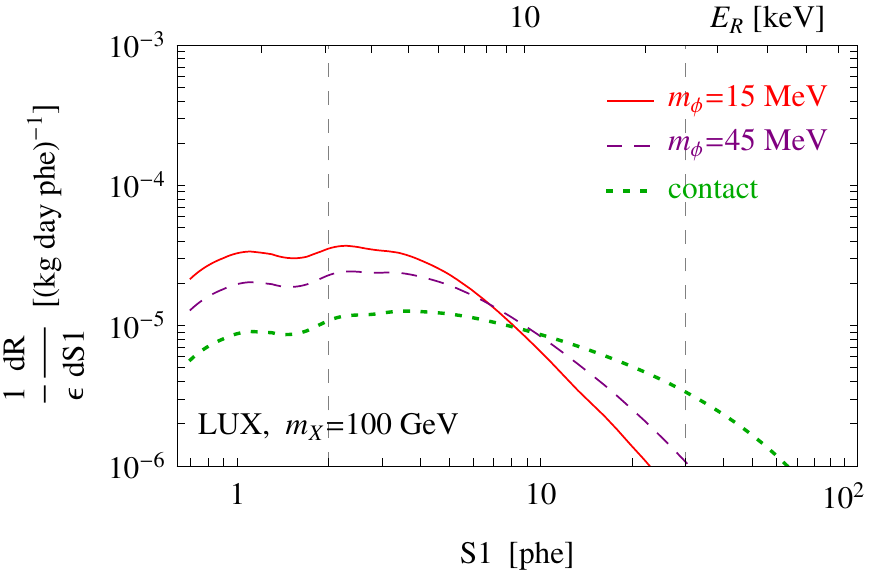}
\includegraphics[width=0.49\textwidth]{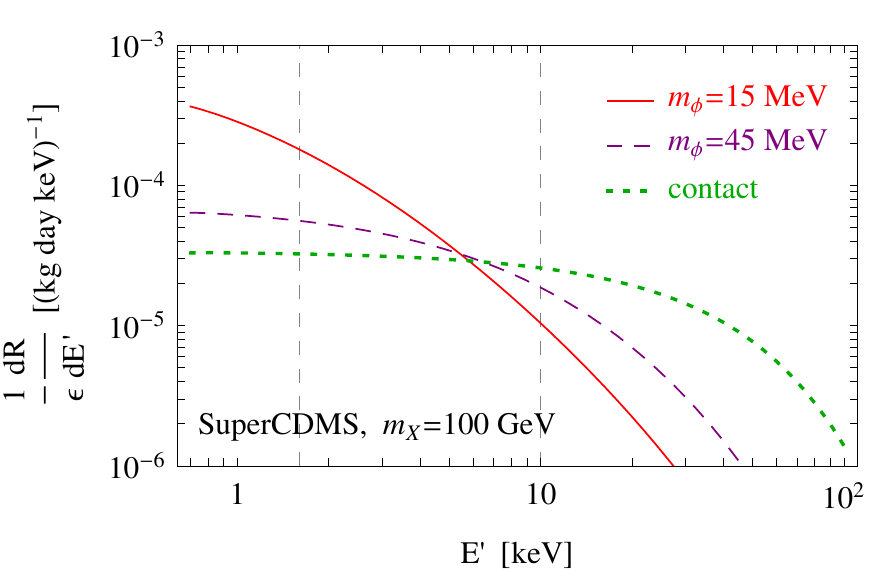}
\caption{\label{figure:spectrumexample}
Differential scattering rates for LUX ({\it left}) and SuperCDMS ({\it right}), for a DM particle with mass $100$ GeV. The solid red line is for our benchmark SIDM model B ($m_\phi = 15$ MeV), the dashed purple line is for a model with three times the mediator mass ($m_\phi = 45$ MeV), and the dotted green line is for the SI model with contact interaction ($m_\phi \gg q$). All curves are normalized to yield the same number of measured events when the measured rate is considered. {\it Top:} theoretical scattering rate as a function of the recoil energy $\ER$. {\it Bottom:} theoretical scattering rate integrated with the resolution function, or in other words, the measured rate prior to including the effect of the experimental efficiency $\epsilon$, as a function of the detected signal ($S1$ for LUX and $\Ed$ for SuperCDMS). Also plotted is the range in detected signal probed by the experiments (dashed vertical lines). For LUX, we also provide on the top axis the average recoil energy $\ER$ corresponding to the detected signal $S1$ in photoelectrons.}
\end{figure}

The quantity that is ultimately measured by the experiments is the differential rate in primary scintillation light/photoelectrons for LUX, $\ud R / \ud S1$ in \Eq{diffrateLUX}, or in detected energy for SuperCDMS, $\ud R / \ud \Ed$ in \Eq{diffrateSuperCDMS}. Therefore, it makes sense to compare these rates for our different DM models, rather than the theoretical recoil rate $\ud R / \ud \ER \equiv \sum_T \ud R_T / \ud \ER$. To understand how the experiment-dependent effects, as resolution and efficiency, affect the spectrum when the rate in $S1$ and $\Ed$ is computed, we show in Fig.~\ref{figure:spectrumexample} the theoretical spectrum ({\it top}) and how it changes after the detector resolution has been taken into account ({\it bottom}), corresponding to neglecting the experimental efficiency $\epsilon$ in the measured rate. We choose our benchmark model B for this illustration, with $m_X = 100$ GeV and $m_\phi = 15$ MeV (solid red line); the model with three times the mediator mass, $m_\phi = 45$ MeV (dashed purple line) is plotted along, together with the SI contact interaction (dotted green line). While theoretical and measured spectra are almost identical for SuperCDMS (before considering $\epsilon$) due to its high resolution, the effect of adding the detector resolution is significant for LUX, in particular in the low $\ER$ region. This is due to the involved process of conversion of the nuclear recoil energy into a signal in noble liquid detectors (and possibly due to the conservative $3$ keV cut in the integral in \Eq{diffrateLUX}); our treatment, while lacking a detailed description of such process, already captures some of the modifications occurring to the spectrum. Also plotted in the bottom panels of Fig.~\ref{figure:spectrumexample} is the $S1$ range of LUX and the detected energy range of SuperCDMS (dashed vertical lines). For LUX, we also include on the top horizontal axis the $\ER$ scale corresponding in average to the $S1$ values on the bottom axis, defined by $\ER = \nu^{-1}(S1)$ (see footnote \ref{nu^-1}).

From Fig.~\ref{figure:spectrumexample} it can be noted that the SIDM model has a spectrum that is peaked towards low nuclear recoil energy, compared to the SI contact interaction model, as expected from the long-range nature of the interaction. For its contact nature, instead, the SI rate is virtually energy independent, if not for the drop at high energies due to the nuclear form factor, which parameterizes the loss of coherence experienced when the projectile (in this case the DM particle) stops `seeing' the nucleus as a whole and starts resolving its internal structure. As anticipated, the model with three times the mediator mass as the SIDM model has a spectrum in between these two cases, and all three spectra are very similar at high energy, when the mediator mass becomes negligible.

\begin{figure}[t!]
\vspace*{-12 mm}
\centering
\includegraphics[width=0.49\textwidth]{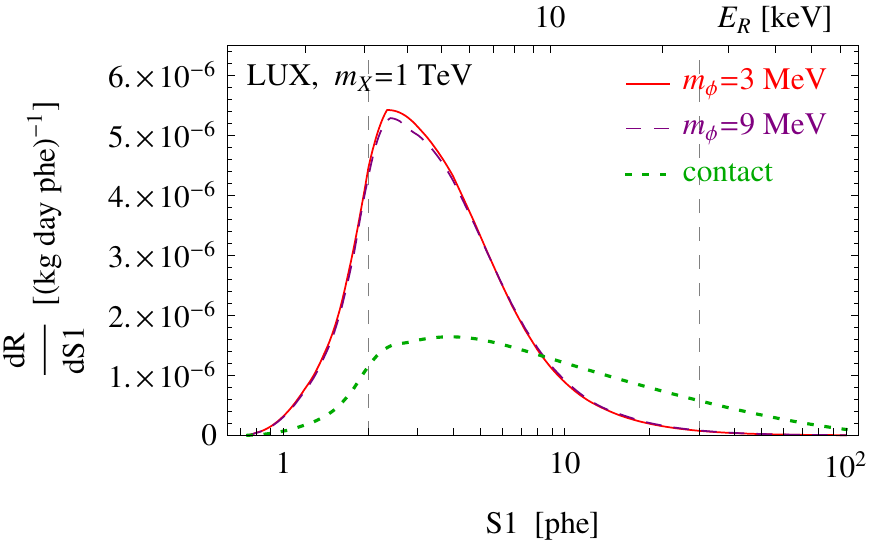}
\includegraphics[width=0.49\textwidth]{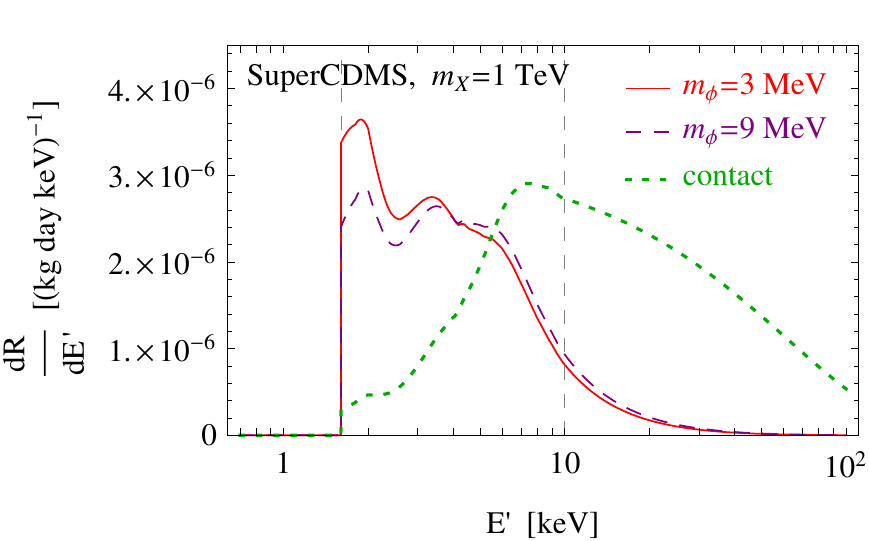}
\includegraphics[width=0.49\textwidth]{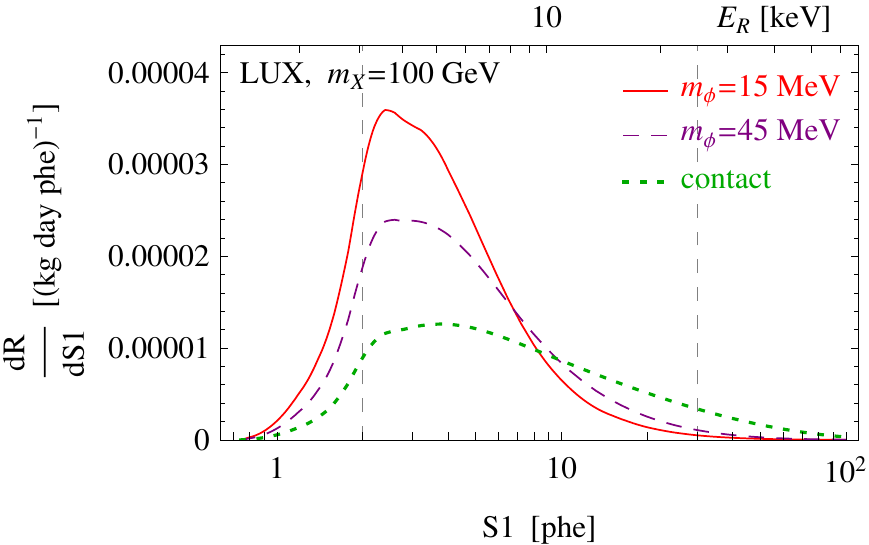}
\includegraphics[width=0.49\textwidth]{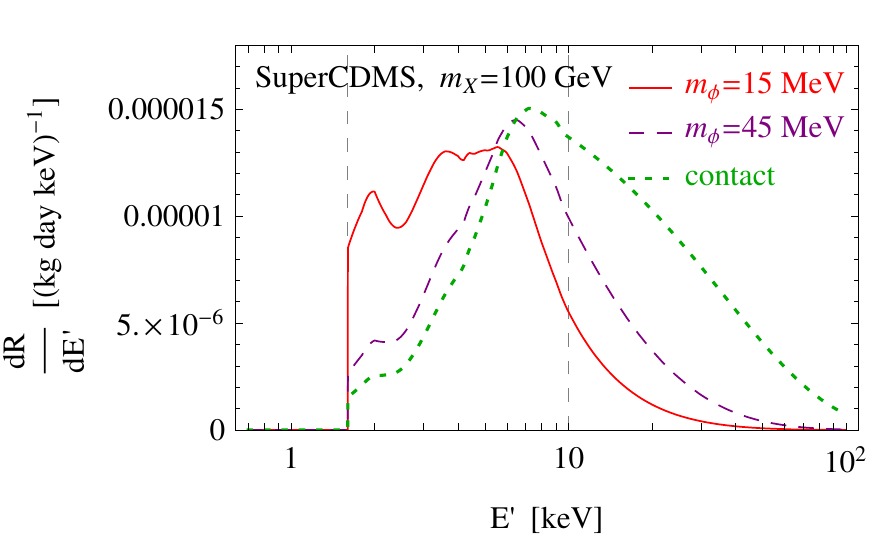}
\includegraphics[width=0.49\textwidth]{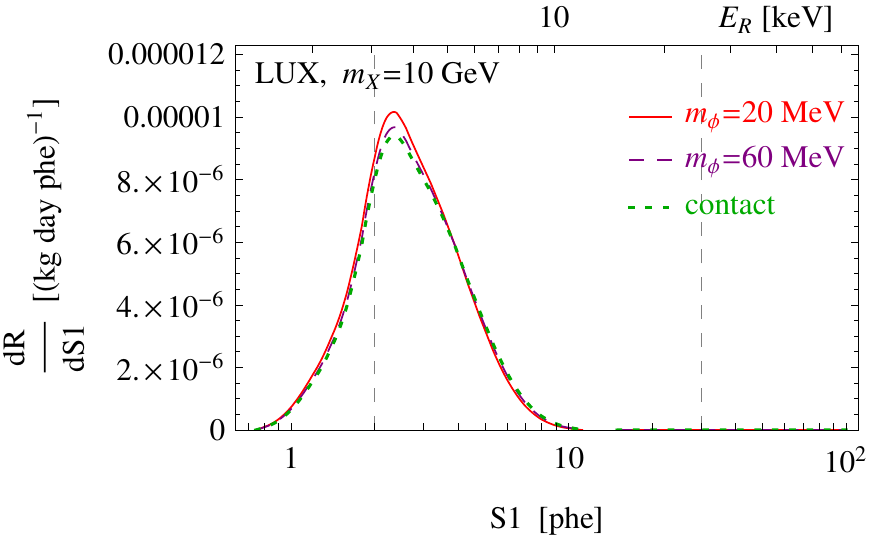}
\includegraphics[width=0.49\textwidth]{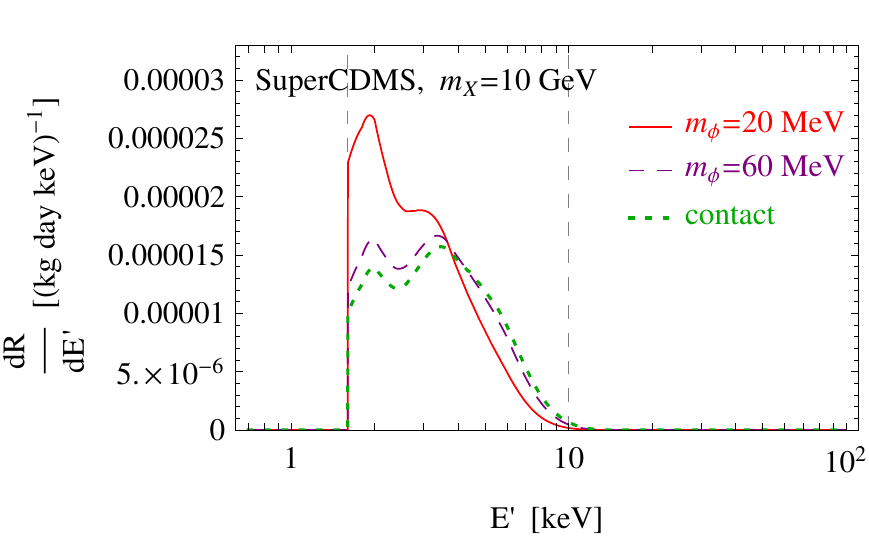}
\rule{0.49\textwidth}{0mm}
\includegraphics[width=0.49\textwidth]{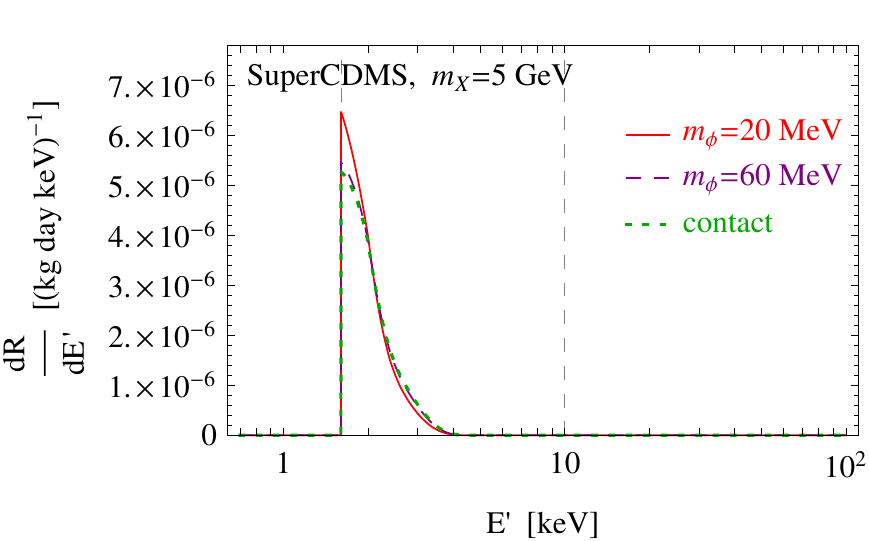}
\caption{\label{fig:spectrum}
Measured rates at LUX ({\it left}) and SuperCDMS ({\it right}), for different DM masses. For each of our benchmark SIDM models (solid red line), a model with three times the mediator mass (dashed purple line) and a SI model with contact interaction (dotted green line) are also considered. The spectra are normalized to have the same area within the signal range, enclosed by the two vertical dashed lines. For LUX, we also provide on the top axis the average recoil energy $\ER$ corresponding to the detected signal $S1$ in photoelectrons. Notice that a $5$ GeV DM particle is below threshold for LUX, and therefore the measured rate is zero.}
\end{figure}

Fig.~\ref{fig:spectrum} shows the measured event spectrum at LUX ({\it left}) and SuperCDMS ({\it right}) for all SIDM benchmark models, each compared to the model with three times the mediator mass and to the SI model with same DM mass. With respect to Fig.~\ref{figure:spectrumexample}, here we include both detector resolution and efficiency. The dashed vertical lines enclose the range in detected signal probed by the experiment. The three curves in each plot are normalized to have the same area within the signal range (same number of measured events), exactly as in Fig.~\ref{figure:spectrumexample}. The experimental efficiency has the general effect of reducing the overall rate; moreover, it is smaller at low energies, as could be guessed from a comparison with Fig.~\ref{figure:spectrumexample}. SuperCDMS's efficiency is taken to vanish below the $1.6$ keV threshold, hence the rate is zero below that energy. LUX's efficiency instead goes to zero smoothly at small values of $S1$. The high energy behavior of the spectrum is dictated by the form factor, which depends solely on the target nucleus and its recoil energy, for $m_X = 1$ TeV and $100$ GeV, and in fact the exponential tails of the differential rate are similar in these two cases. The velocity integral, which is almost constant within the LUX and SuperCDMS signal region for these DM masses, moves to lower energies when decreasing the DM mass, and for $m_X$ as small as $5$ or $10$ GeV it provides the main source of suppression of the rate at high energies (see Fig.~\ref{fig:eta}). For this reason the high energy tail of the differential rate moves to lower energies when decreasing the DM mass, while the low energy tail is fixed by the efficiency function. For light enough DM, the velocity integral vanishes at energies below the experimental threshold and therefore there is no detectable signal, unless the energy threshold is lowered. This is what happens with a $5$ GeV DM particle at LUX, as can be seen in Fig.~\ref{fig:eta}, for which the measured rate is zero.

The lighter mediators move the recoil spectrum to lower energies as expected from \Eq{sigma_XT}. A more quantitative understanding of the transition between the long-range regime and the contact regime can be obtained by looking at Table \ref{tb:models}. For a $1$ TeV DM particle as in our benchmark model A, the typical momentum transfer is $127$ MeV for LUX and $74$ MeV for SuperCDMS. Therefore, the DM-nucleus scattering for benchmark model A, with a $3$ MeV mediator, is deeply in the long-range regime $m_\phi \ll q$ for both experiments and the corresponding recoil spectrum is clearly different from the spectrum of the SI model with contact interaction. The same can be concluded for the model with a $9$ MeV mediator. Scattering for the benchmark model B, with $m_X = 100$ GeV and $m_\phi = 15$ MeV, again occurs in the long-range regime, while for a $45$ MeV mediator we have $m_\phi \approx q$ and therefore the scattering occurs in between the pure long-range and the contact regimes. The $m_\phi \approx q$ regime is also relevant for scattering in SuperCDMS for the benchmark model C, with a $10$ GeV DM particle and $m_\phi = 20$ MeV, while a $60$ MeV mediator is much closer to the contact interaction model. In LUX this model is difficult to distinguish from WIMP models with contact interaction, because the scattering occurs close to the energy threshold of the detector. In fact, because of the small DM mass, the high energy tail of the spectrum (dominated by the exponential fall-off of the velocity integral $\eta_0$ at large $\vmin$) gets very close in energy to the low energy tail (dominated by the efficiency function $\epsilon$).  Neither $\eta_0$ nor $\epsilon$ depend on the mediator mass, and the only difference between the various models can be seen at the peak of the recoil spectrum, where the $(q^2 + m_\phi^2)^{-2}$ dependence of the cross section has a tiny effect. The same behavior can be seen for benchmark model D ($m_X = 5$ GeV) in SuperCDMS.

\begin{figure}[t!]
\centering
\includegraphics[width=0.49\textwidth]{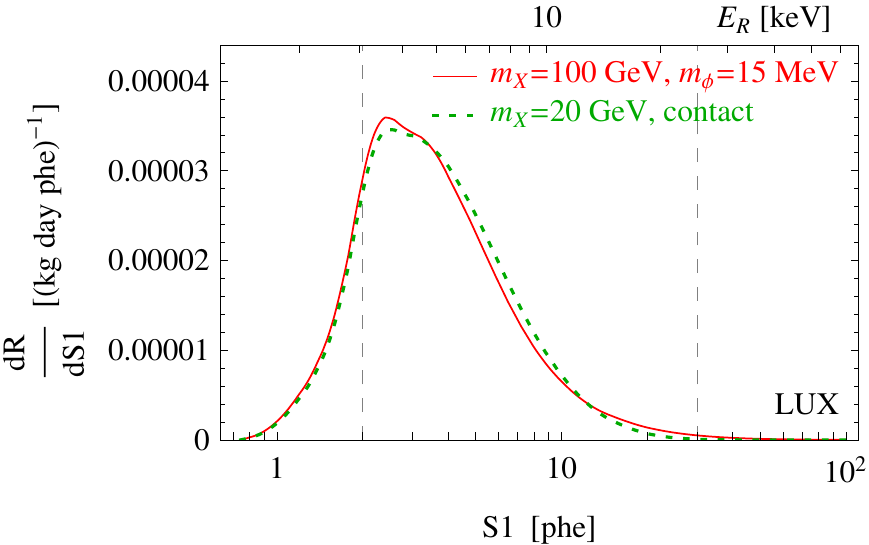}
\includegraphics[width=0.49\textwidth]{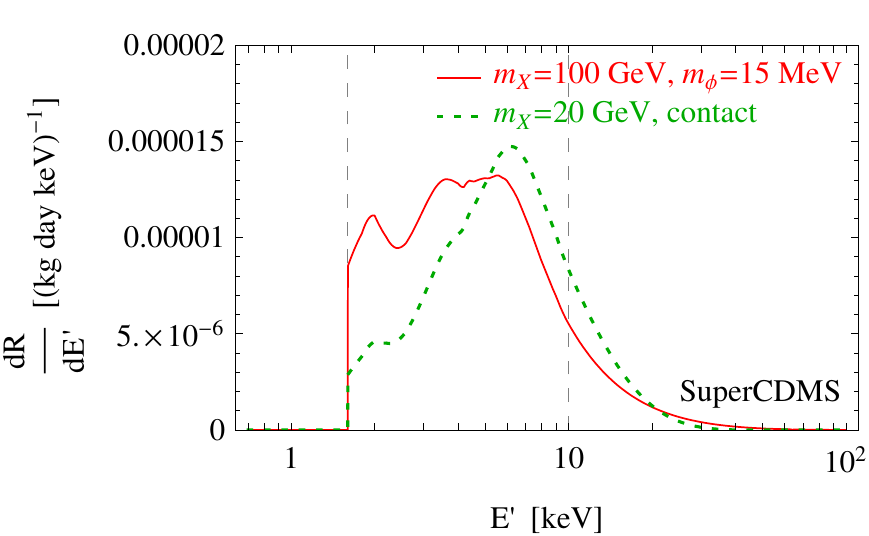}
\includegraphics[width=0.49\textwidth]{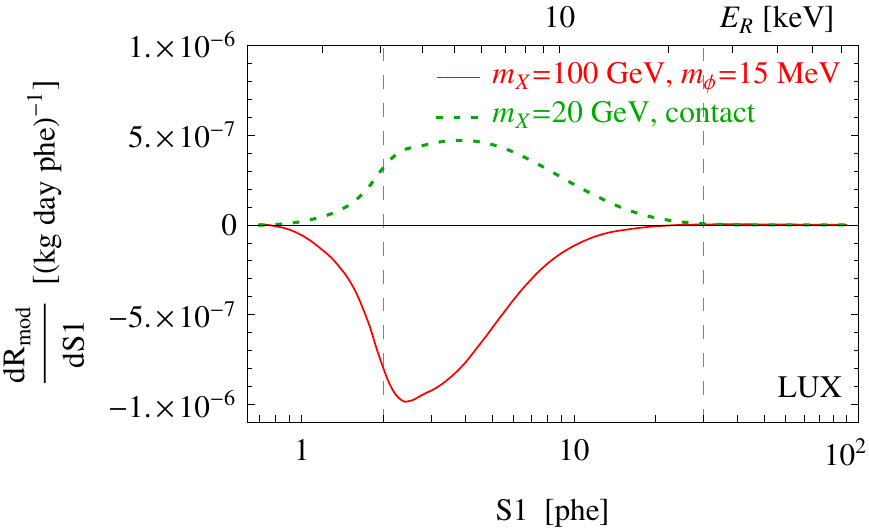}
\includegraphics[width=0.49\textwidth]{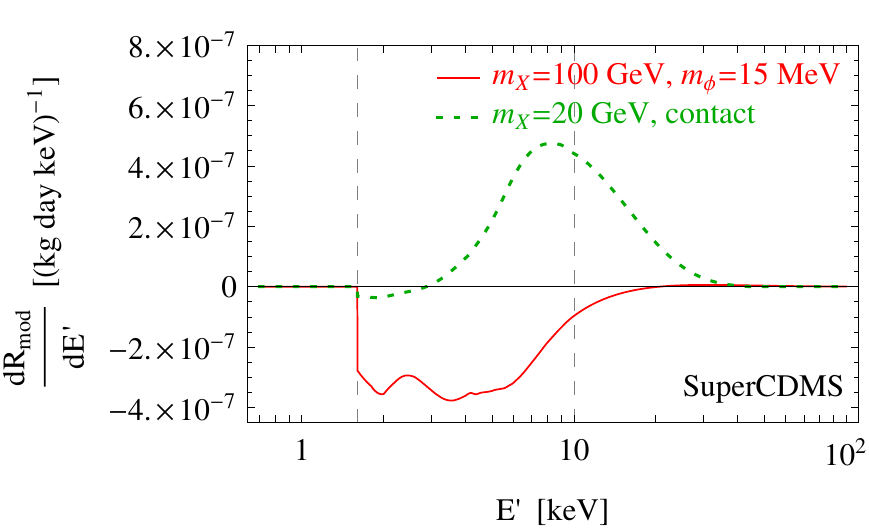}
\caption{\label{fig:comparison}
Differential scattering rates for LUX ({\it left}) and SuperCDMS ({\it right}). Both the unmodulated ({\it top}) and modulated ({\it bottom}) components of the differential rate are shown. The solid red line is for our benchmark SIDM model B ($m_X = 100$ GeV, $m_\phi = 15$ MeV), while the dotted green line is for a $20$ GeV DM particle with contact interactions. All curves are normalized to have the same area between the two dashed vertical lines, indicating the signal range probed by the experiment.}
\end{figure}

We have argued that long-range and contact interactions can be distinguished by the different spectrum at direct detection experiments, since the recoil spectrum is flatter at low energies for contact interactions. However, since only energies above a certain threshold are accessible at the experiments, a light WIMP could fake the steeper SIDM spectrum. In fact, as discussed above, the flat part of the spectrum for a light WIMP lies below threshold for both LUX and SuperCDMS, so that these experiments can only observe the tail due to the exponential drop of the velocity integral. This is the reason why the assumption of a heavy mediator favors lighter DM in the analysis of direct detection data, while assuming a long-range interaction points toward heavier DM, as noticed in Ref.~\cite{Li:2014vza}. From a visual inspection of the recoil spectra we ascertain that $10$--$20$ GeV WIMPs can mimic the SIDM spectra, depending on the target material but almost independently of the DM and mediator mass in the SIDM model. To assess the ability of the experiments to distinguish the two signals, we show in \Fig{fig:comparison} the spectrum of a $20$ GeV DM particle with contact interaction and compare it with the spectrum from our benchmark model B ($m_X = 100$ GeV, $m_\phi = 15$ MeV), for both LUX ({\it left}) and SuperCDMS ({\it right}). It is evident from the upper panels that the WIMP and SIDM spectra at LUX are very similar. However, the two spectra could be distinguished by an experiment employing a different target, such as SuperCDMS. Furthermore, the two signals have very different modulated spectra, as shown in the bottom panels of \Fig{fig:comparison}. The modulated spectrum of the WIMP model is almost entirely positive above threshold, as is typical for light DM particles, while the SIDM spectrum is partially negative because of the large DM mass. The shape of the modulated spectrum is explained in detail in the next section for a sodium iodide detector like DAMA. Since the modulated rate is about $30$--$60$ times smaller than the unmodulated rate, we expect that hundreds of events will be needed in order to distinguish SIDM from WIMPs using modulation signals.

\medskip

Our results for this section can be summarized as follows. Despite the limitations imposed by detector resolution and efficiency, direct detection experiments such as LUX and SuperCDMS can potentially distinguish SIDM and WIMP recoil spectra, provided the DM mass is heavy enough that the scattering does not occur too close to the experimental threshold. While light WIMPs may fake a SIDM signal at a direct detection experiment, the degeneracy between the two spectra could be lifted either by a second experiment employing a different target, or by observing the modulated part of the spectrum. While LUX has a larger sensitivity to heavy DM and a much higher exposure, SuperCDMS is more sensitive to low-mass DM particles because of its lighter target and low energy threshold.

\subsection{Modulation signal}
Together with the average rate at LUX and SuperCDMS, we also study the prospects to distinguish SIDM from WIMPs using the annually modulated part of the rate. While the modulated rate is usually much smaller than the average rate and therefore more difficult to detect, it is a cleaner signature of DM detection because the modulation phase has to be consistent with Earth's motion through the DM distribution.

\begin{figure}
\centering
\includegraphics[width=0.49\textwidth]{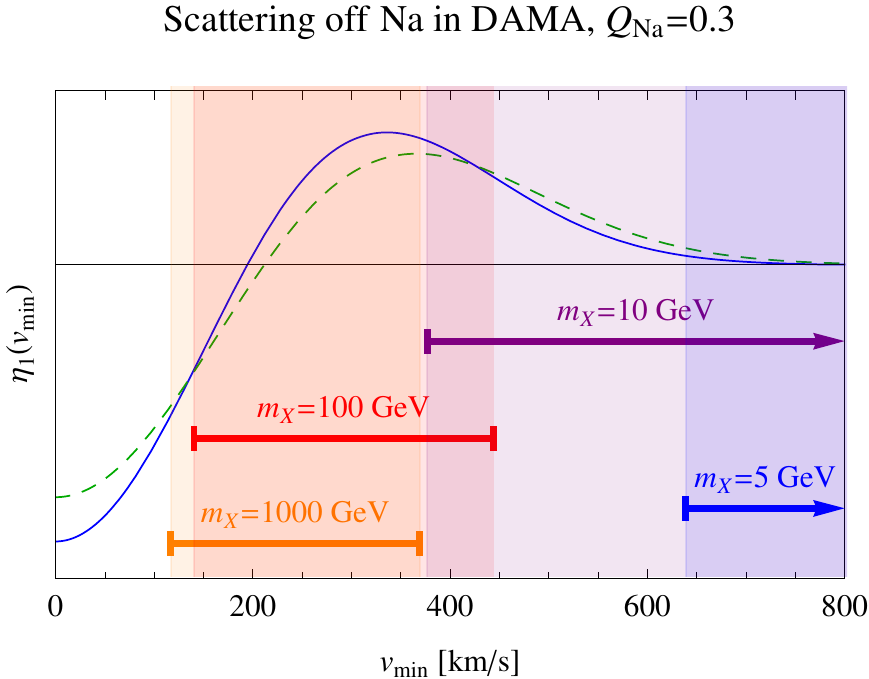}
\includegraphics[width=0.49\textwidth]{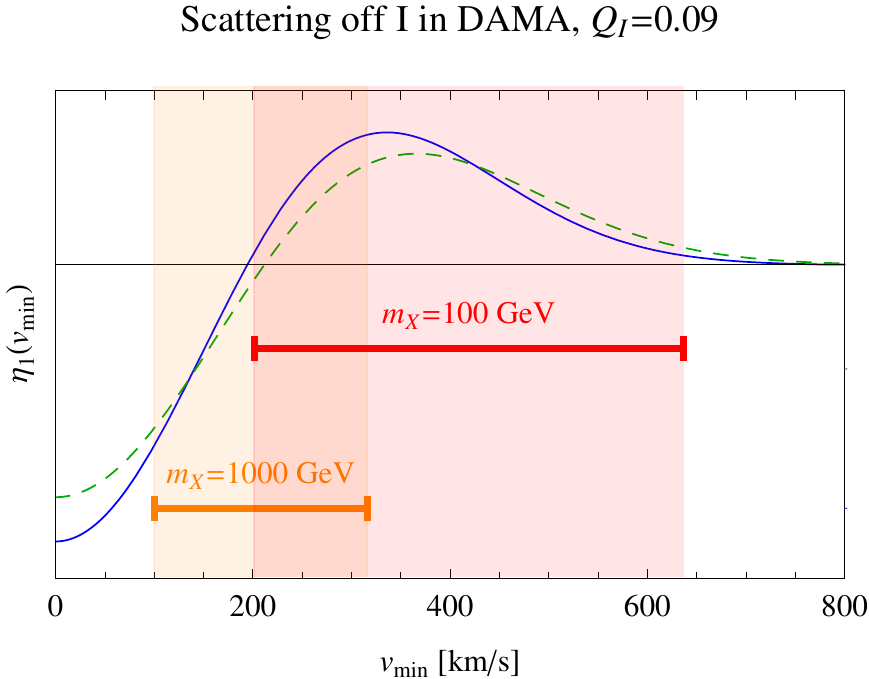}

\vspace{3mm}

\includegraphics[width=0.49\textwidth]{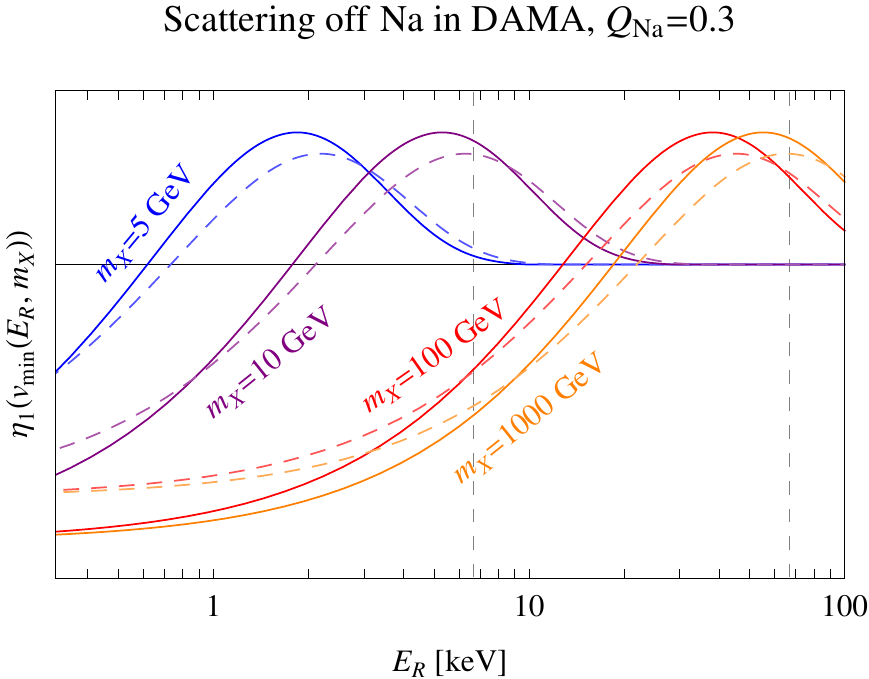}
\includegraphics[width=0.49\textwidth]{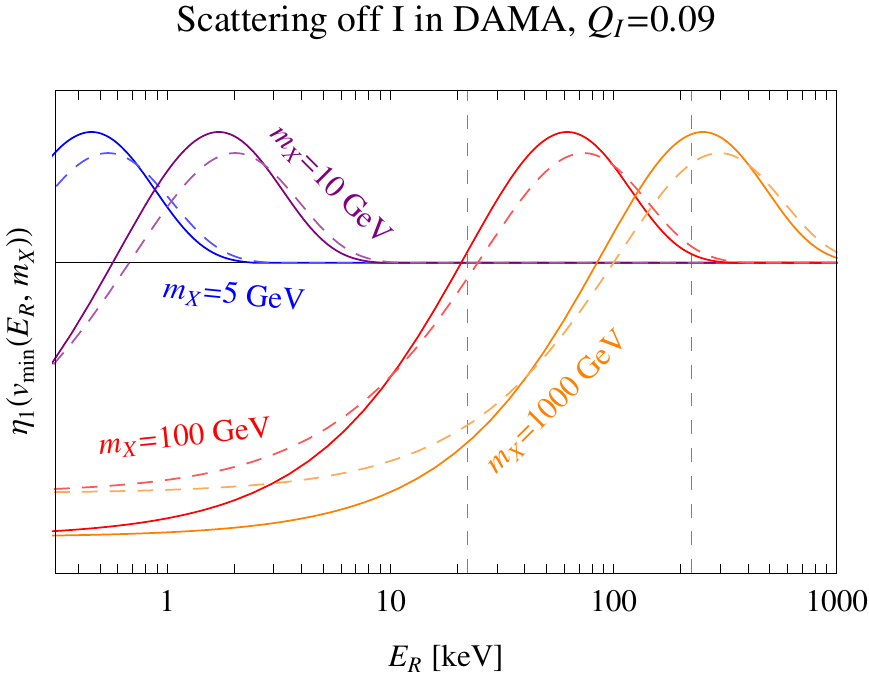}
\caption{Modulated velocity integral in the SHM. Colors and line styles are as in \Fig{fig:eta}. The ranges in nuclear recoil energy shown are $[6.7 \text{ keV}, 66.7 \text{ keV}]$ for scattering off sodium and $[22.2 \text{ keV}, 222.2 \text{ keV}]$ for scattering off iodine.}
\label{fig:eta'}
\end{figure}

The DAMA Collaboration has claimed a $9.3 \sigma$ evidence for modulation in their data~\cite{Bernabei:2013xsa}, although this result is in strong tension with experiments measuring no signal. Here we take DAMA as an example to study the annual modulation signal. Our results will apply straightforwardly to similar future experiments such as KIMS-NaI, ANAIS, DM-Ice17, and SABRE (see \eg\cite{Cooley:2014aya} and references therein). Our analysis follows that in the previous section, with the modulated rate given by \Eq{simple rate} with $\eta_0$ substituted by $\eta_1$. Analogous to \Fig{fig:eta}, we plot in \Fig{fig:eta'} the functions $\eta_1(\vmin)$ in $\vmin$-space ({\it top}) and $\eta_1(\vmin(\ER, m_X))$ in $\ER$-space ({\it bottom}), for the two target nuclei used in the experiment, sodium ({\it left}) and iodine ({\it right}). The solid lines are for our fiducial SHM parameters, while the dashed lines are for the larger values reported in the previous section. The horizontal black line indicates the zero of $\eta_1$; positive values indicate that the modulation phase $t_0$ in \Eq{rateexpansion} corresponds to the time of maximum of the signal at $\vmin \gtrsim 200$ km/s, while negative values indicate that $t_0$ becomes the time of minimum rate for $\vmin \lesssim 200$ km/s. The experimental range in detected energy, $\Ed \in [2 \text{ keV}, 20 \text{ keV}]$, is mapped onto a different range in $\ER$ for each target because of the different quenching factors for sodium and iodine, $Q_\text{Na} = 0.3$ and $Q_\text{I} = 0.09$, respectively (the quenching factor being the average fraction of recoil energy that is recorded as detected energy). The corresponding $\vmin$ ranges are displayed as colored bands for each choice of DM mass in our four benchmark models in the top panels, and the $\ER$ range is delimited by the two vertical dashed lines in the bottom panels. For $m_X = 10$ GeV and $m_X = 5$ GeV, the signal window continues to the right of the plotted range for sodium, and lies completely outside of the plotted range for iodine. From the figure it is apparent that scattering off sodium is marginally sensitive to a $5$ GeV DM, while it is fully sensitive to $m_X = 10, 100,$ and $1000$ GeV. On the other hand, because of its larger mass, scattering off iodine is completely below threshold for light DM, $m_X = 5$ and $10$ GeV.

As mentioned in Sec.~\ref{sec:direct detection}, our form of the modulation signal relies on the DM distribution being isotropic in its rest frame. DM substructure, as \eg dark disks and streams, will in general modify the modulation amplitude and its phase. We are also neglecting the anisotropy induced by the Sun's gravitational potential, which modifies the phase of the modulation below $\sim 200$ km/s~\cite{Lee:2013wza, Bozorgnia:2014dqa}.

\begin{figure}[t!]
\centering
\includegraphics[width=0.49\textwidth]{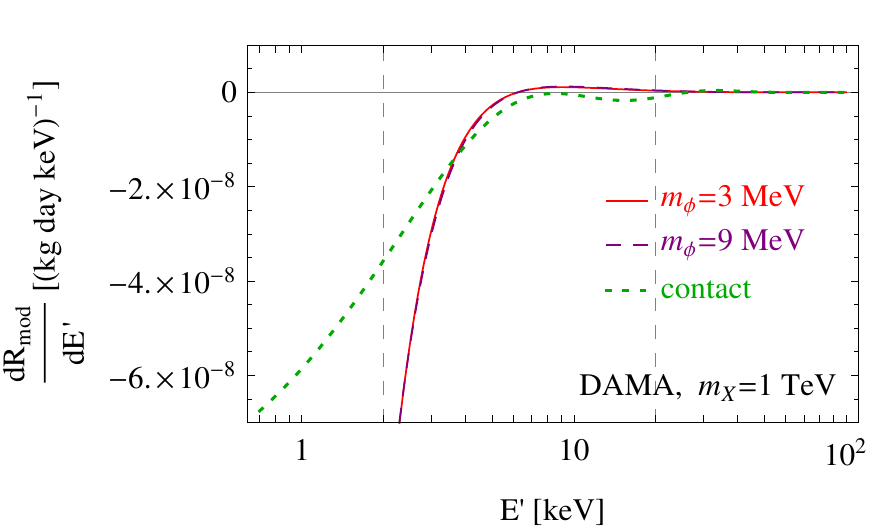}
\includegraphics[width=0.49\textwidth]{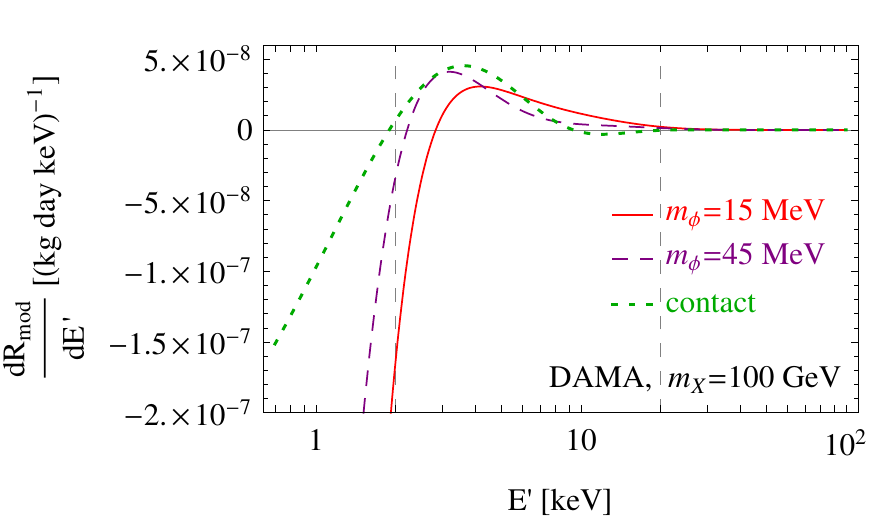}
\includegraphics[width=0.49\textwidth]{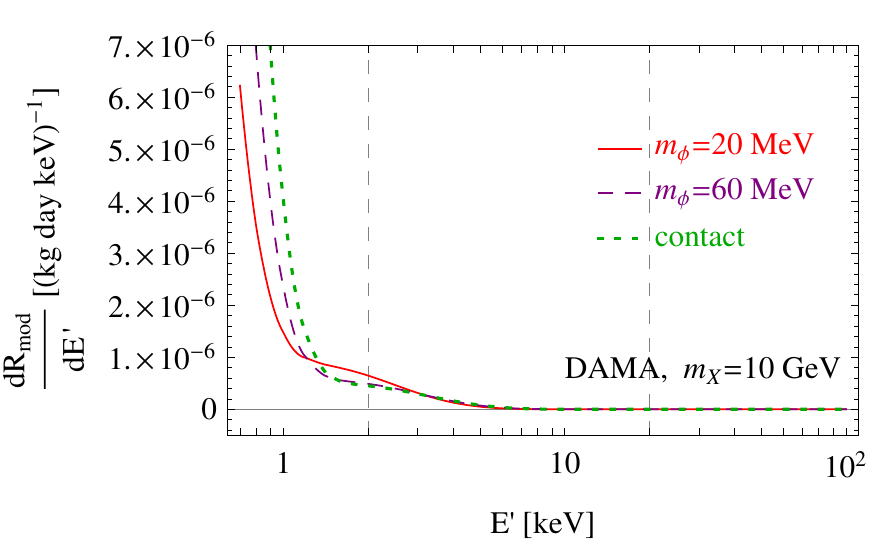}
\includegraphics[width=0.49\textwidth]{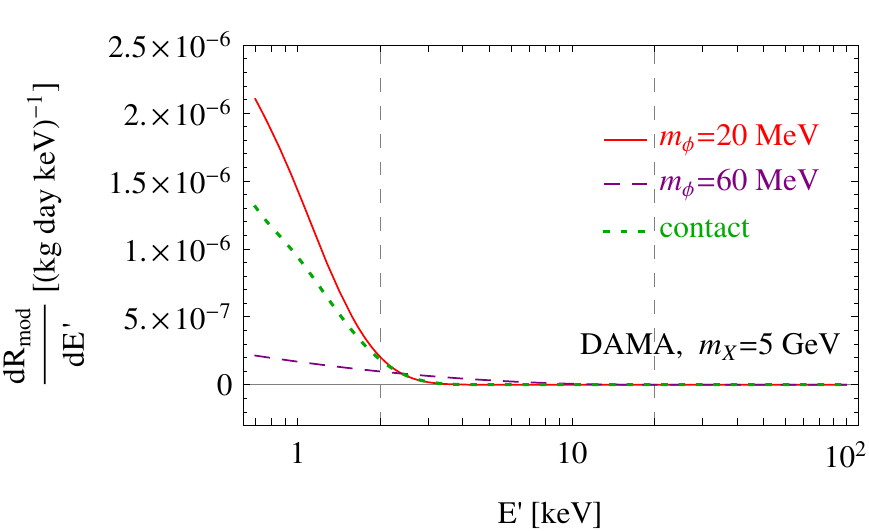}
\caption{\label{fig:DAMA}
Measured modulated rate at DAMA for different DM masses. For each of our benchmark SIDM models (solid red line), a model with three times the mediator mass (dashed purple line) and a SI model with contact interaction (dotted green line) are also considered. The spectra are normalized to have the same area within the signal range, enclosed by the two vertical dashed lines.}
\end{figure}

In \Fig{fig:DAMA} we plot the modulated spectrum for DAMA, in analogy with the unmodulated spectrum for LUX and SuperCDMS in \Fig{fig:spectrum}. The modulated spectrum is shown for each of our four benchmark models (solid red lines), together with the additional model with three times the mediator mass (dashed purple lines) and the WIMP model with SI contact interaction (dotted green lines). The three spectra are normalized such that the total modulated rate in the signal window of the experiment is the same (same number of events after subtracting the unmodulated rate). The plots show that it may be possible to distinguish SIDM from WIMPs with the modulated component of the rate, especially if the experimental sensitivity extends to low energies.

\section{Conclusions}
\label{sec:conclusions}
DM self-interactions are a promising avenue to solve the small-scale puzzles in galaxy formation. If the self-interations are modeled as a Yukawa potential, the mediator mass required to generate large enough self-interactions to affect the internal structure of galaxies should be generically below $100$ MeV. In order to avoid overclosing the universe, the mediator must be unstable. We have assumed that the mediator decays to lighter SM particles, which opens up the possibility of searching for SIDM particles in direct detection experiments.

In this paper we have studied SIDM direct detection in detail. While we have focused on SIDM benchmarks, our work is applicable to all DM models with  long-range interactions with nuclei. We have shown that DM direct detection experiments are remarkably sensitive to the SIDM parameter space, even if the coupling between the two sectors is extremely feeble. For example, LUX has excluded all the favored SIDM region with DM heavier than $7$ GeV if the kinetic mixing parameter is larger than $10^{-9}$. Models with DM particle masses down to $3$ GeV can be excluded by SuperCDMS for values of the kinetic mixing parameter an order of magnitude larger.

In contrast to the point-like interaction in usual WIMP-nucleus scattering, SIDM interacts with nuclei through a long-range force, which leads to a measurably different signal spectrum. When the mediator mass is comparable to the momentum transfer of nuclear recoils, the SIDM spectrum is peaked more towards low recoil energy compared to usual WIMPs, and can be potentially tested by direct detection experiments such as LUX, SuperCDMS, and sodium iodide experiments like DAMA. Our analysis shows that we may be able to distinguish long-range (SIDM) and contact interaction spectra with detections in experiments with different targets or by measuring both the time-average and the modulated rate.

\section*{Acknowledgments} 
\noindent
E.D.N.~acknowledges partial support from the U.~S.~Department of Energy under Grant No.~DE-SC0009937. M.K.~is supported by NSF Grant No.~PHY-1214648 and No.~PHY-1316792. H.-B.Y.~is supported in part by the U.~S.~Department of Energy under Grant No.~DE-SC0008541.

\bibliography{sidmbib}

\end{document}